\documentclass[11pt,a4paper]{article}

\usepackage[margin=1in]{geometry}

\usepackage{amsmath,amsfonts,amssymb,amsthm}
\usepackage{array}

\usepackage{graphicx}
\usepackage{subcaption}
\usepackage{algorithm}
\usepackage{algorithmic}

\usepackage{textcomp}
\usepackage{url}
\usepackage{xcolor}
\usepackage{hyperref}
\usepackage{balance}

\newcommand{\SO}{\mathsf{SO}}
\newcommand{\R}{\mathbb{R}}

\newcommand{\F}{\text{F}}
\newcommand{\tr}{\textnormal{tr}}
\newcommand{\hf}{F}
\newcommand{\hg}{G}
\newcommand{\hy}{Y}
\newcommand{\he}{\mathcal{E}}
\newcommand{\U}{\mathrm{U}}
\newcommand{\Fpop}{\breve{F}}

\newtheorem{thm}{Theorem}[section]
\newtheorem{lemma}[thm]{Lemma}

\newtheorem{remark}[thm]{Remark}

\DeclareMathOperator*{\argmax}{arg\,max}

\title{A Fast Approximate Maximum Likelihood Estimator for Low SNR Multi-Reference Alignment}

\author{
	Shay Kreymer, Amnon Balanov, and Tamir Bendory\thanks{
		Shay Kreymer is supported by the TAD Excellence Program for Doctoral Students in Artificial Intelligence and Data Science. Tamir Bendory is supported in part by BSF under Grant 2020159, in part by NSF-BSF under Grant 2019752, and in part by ISF under Grant 1924/21. Corresponding author: Shay Kreymer (shaykreymer@mail.tau.ac.il).
	}\\[3pt]
	School of Electrical and Computer Engineering, Tel Aviv University, Tel Aviv  6997801, Israel
}

\date{}

\begin{document}
	
	\maketitle
	
	\begin{abstract}	
		Motivated by single-particle cryo-electron microscopy, multi-reference alignment (MRA) models the task of recovering an unknown signal from multiple noisy observations corrupted by random rotations. The standard approach for computing the maximum likelihood estimator (MLE) is the expectation-maximization (EM) algorithm; however, it often becomes computationally prohibitive, particularly in low signal-to-noise ratio (SNR) settings. We introduce a fast approximate MLE for MRA over the special orthogonal groups $\mathrm{SO}(2)$ and $\mathrm{SO}(3)$ in the low-SNR regime. A low-SNR Taylor expansion of the likelihood reveals a closed-form, non-iterative approximate MLE. We show that this approach yields a consistent estimator in the low-SNR limit and requires substantially lower computational complexity than both EM and invariant-based alternatives. Numerical experiments generated from molecular volumes show that the proposed method provides a favorable accuracy-runtime trade-off, especially in challenging low-SNR regimes, and can serve as an effective initialization for EM.
\end{abstract}

\section{Introduction}
\label{sec:intro}
Multi-reference alignment (MRA) is the problem of estimating a signal from multiple noisy observations, each transformed by an unknown group action.
This paper focuses on MRA over the special orthogonal group~$\SO(3)$, in which the goal is to recover a three-dimensional signal from noisy observations that have been independently rotated by unknown elements of~$\SO(3)$. This model is closely related to subtomogram averaging in cryo-electron tomography (cryo-ET), where three-dimensional subtomograms extracted from tomographic reconstructions must be aligned and averaged despite unknown orientations~\cite{chen2019complete,zivanov2022bayesian}. It also provides a simplified mathematical setting for studying algorithmic challenges arising in single-particle cryo-electron microscopy (cryo-EM), where a three-dimensional molecular structure is reconstructed from noisy two-dimensional projection images acquired at unknown viewing directions~\cite{nogales2016development,singer2020computational,bendory2020single, subramaniam2026cryo}. Although cryo-EM additionally involves a tomographic projection operator and other modeling complications, MRA over~$\SO(3)$ retains the central difficulty of estimating a signal in the presence of unknown latent rotations and severe noise~\cite{bandeira2023estimation,bendory2025orbit}.

The principal challenge in MRA lies in the low-SNR regime, where individual observations are so heavily corrupted by noise that estimating the corresponding group actions is statistically infeasible. Remarkably, despite the impossibility of recovering the individual transformations, consistent estimation of the underlying signal is still achievable, provided that the number of observations scales faster than~$\sigma^6$, where~$\sigma$ is the standard deviation of the noise~\cite{perry2019sample}. A statistically natural approach to MRA is maximum likelihood estimation (MLE), typically implemented by expectation-maximization~(EM)~\cite{dempster1977maximum}. EM treats the unknown rotations as latent variables, estimating the signal by maximizing the expected log-likelihood with respect to the posterior distribution of these rotations. However, EM suffers from two major drawbacks. First, the likelihood landscape is non-convex, and EM is prone to getting trapped in spurious local minima unless initialized with a high-quality estimate~\cite{fan2023likelihood,fan2024maximum}. Second, EM is computationally expensive, as it requires multiple iterations over the entire dataset. In particular, as the SNR decreases, the convergence of EM can become extremely slow, even near the solution~\cite{balanov2025expectation}.

As an alternative to EM, it was suggested to avoid the latent rotations altogether by using invariant features, such as moments and bispectra~\cite{bendory2017bispectrum, abbe2018estimation, bendory2025orbit}. These methods provide consistent estimators under suitable conditions, and require a single pass over the whole dataset. In high-dimensional settings, however, invariant-feature recovery requires solving large systems of  polynomial equations. This becomes especially computationally challenging for $\SO(3)$ models and for more realistic imaging problems motivated by cryo-EM.

The connection between likelihood-based estimation and invariant moments in the low-SNR regime has been studied in several recent works~\cite{katsevich2022likelihood, fan2023likelihood, fan2024maximum,balanov2026generalized}. These works provide an initial theoretical explanation for why low-order invariants govern estimation in the high-noise limit. The present paper builds on this viewpoint in an algorithmic direction and derives explicit closed-form MLE estimators for MRA over~$\SO(2)$ and~$\SO(3)$.
In particular, we introduce a computationally efficient approximation to the MLE that avoids iterative refinement. The proposed algorithm is statistically consistent in the low-SNR regime. Leveraging the low-SNR assumption, we expand the marginalized likelihood to derive a sequential ``frequency marching'' estimator that propagates the unknown degrees of freedom from lower to higher frequencies based on data-driven averages of the observations. To demonstrate the leading principles of the method, we begin by deriving the algorithm for MRA over the~$\SO(2)$ group. In this case, these degrees of freedom are scalar Fourier phases, while in the $\SO(3)$ case they are unitary matrices associated with spherical-harmonic coefficients. Our primary contribution is the derivation of this analytic approximation of the likelihood function of MRA over~$\SO(2)$ and~$\SO(3)$ in the low-SNR regime, which yields a non-iterative, computationally efficient closed-form algorithm.

The resulting estimator is closely related to bispectrum inversion, but is derived from the likelihood rather than from explicitly constructed invariants. Whereas bispectrum methods form and invert third-order invariant tensors, we instead view the marginalized log-likelihood as a function of the single unresolved degree of freedom of the next frequency band, holding the lower-frequency components fixed. Our consistency analysis (Theorems~\ref{thm:consistency_so2} and~\ref{thm:consistency_so3}) shows that the leading-order term of the resulting band update is precisely the third-order rotational invariant underlying bispectrum frequency marching. The proposed method therefore recovers the same low-order information as the bispectrum, but obtains it implicitly from the low-SNR likelihood expansion, avoiding both the iterations of EM and the explicit construction and inversion of large bispectrum systems of equations.

We prove that the proposed estimator is consistent in the low-SNR asymptotic regime as $n,\sigma\to\infty$ with $n=\omega(\sigma^6)$, which is the optimal estimation rate.\footnote{$n=\omega(\sigma^6)$ means that $n/\sigma^{6}\to\infty$ as $n, \sigma \to\infty$.} We further demonstrate numerically that it provides a favorable accuracy-runtime trade-off and can serve as an effective initialization for EM when additional likelihood-based refinement is desired. Thus, this work serves as a foundational step, establishing a concrete algorithmic framework based on low-SNR likelihood expansion. In Section~\ref{sec:discussion}, we briefly discuss how the same principle may extend to broader classes of compact group actions and outline its potential application to the full three-dimensional cryo-EM reconstruction problem.

\section{Fast MLE algorithm for low-SNR MRA over~$\SO(2)$}
Although the main  motivation of this work is MRA over~$\SO(3)$, we begin with the simpler~$\SO(2)$ model to introduce the likelihood-expansion principle. We begin by defining the MRA model over~$\SO(2)$ and the marginalized likelihood. We then derive a first-order approximation of the conditional likelihood in the low-SNR regime, leading to a frequency marching algorithm that recovers the Fourier phases sequentially in closed form. Finally, we discuss the computational complexity of the proposed algorithm relative to two baselines for comparison: the standard EM algorithm~\cite[Section VI]{bendory2017bispectrum} which is the natural likelihood-based approach, and bispectrum inversion~\cite{bendory2017bispectrum}, which is an efficient invariant-feature method for~$\SO(2)$ MRA. 

\subsection{Statistical model} 
We consider a real bandlimited signal on the unit circle:
\begin{equation}
	f(t) = \sum_{k=-L}^L \hf[k] e^{ikt}, \quad t\in[0,2\pi),
\end{equation}
where $L$ is the bandlimit of the signal. Since~$f$ is real, its Fourier transform is conjugate symmetric, i.e., $\hf[-k] = \overline{\hf[k]}$. For an angle~$\theta\in[0,2\pi)$, the rotation operator~$R_\theta \in \SO(2)$ is defined by $(R_\theta f)(t) = f(t - \theta) = \sum_{k=-L}^L \hf[k] e^{-i k \theta} e^{i k t}.$

We consider a set of~$n$ observations,
\begin{equation}
	y_j = R_{\theta_j} f + \varepsilon_j, \quad j = 1, \ldots n,
\end{equation}
where~$\varepsilon_j\stackrel{\mathrm{i.i.d.}}{\sim}\mathcal{N}(0,\sigma^2 I)$, and~$\theta_j \stackrel{\mathrm{i.i.d.}}{\sim} \text{Unif}\left[0, 2 \pi\right)$. The assumption that the distribution over the angles is uniform is unnecessary, and we take it only for simplicity. In the Fourier domain, the observations satisfy
\begin{equation}\label{eq:observations}
	\hy_j[k] = \hf[k]\,e^{-ik\theta_j} + \he_j[k],
\end{equation}
where $\he_j \stackrel{\mathrm{i.i.d.}}{\sim} \mathcal{CN}(0,\sigma^2 I_{L+1})$, also satisfying conjugate symmetry. We note that the signal can be estimated only up to a global rotation.

\subsection{Fast approximate MLE for low SNR MRA over~$\SO(2)$}
The proposed algorithm has two stages. First, it uses the first and second-order empirical moments to estimate the zero-frequency coefficient and the magnitudes of the Fourier coefficients. Then, the missing Fourier phases are recovered sequentially, from lower to higher frequencies, by restricting the marginalized log-likelihood to the phase of the next coefficient while holding the other coefficients fixed. A low-SNR expansion of this restricted log-likelihood yields a closed-form phase update, leading to the fast approximate MLE summarized in Algorithm~\ref{alg:fast_mle}.
\subsubsection{Preliminary estimates}
The first stage estimates the rotation-invariant information carried by the first and second-order moments. Since the zero-frequency component is unaffected by rotations, it can be estimated by averaging:
\begin{equation}
	\label{eq:avg}
	\widehat{\hf}[0] = \frac{1}{n} \sum_{j=1}^n \hy_j[0],
\end{equation}
This estimator is consistent provided that $n=\omega(\sigma^2)$.
For notational simplicity, we denote by $r_k \triangleq |\hf[k]|$ the magnitude of the $k$-th Fourier coefficient. Then, for each nonzero frequency $k=1,\ldots,L$, the magnitude $r_k$ is estimated from the second moment:
\begin{equation}
	\label{eq:magnitudes}
	\hat{r}_k^2 = \max \left\{ {\frac{1}{n} \sum_{j=1}^n |\hy_j[k]|^2 - \sigma^2}, 0 \right\}.
\end{equation}
These estimators are consistent provided that $n=\omega(\sigma^4)$.

At this stage, the zero-frequency coefficient and the magnitudes of the nonzero Fourier coefficients have been estimated, and it remains to recover their phases. Since the signal is real-valued, its Fourier coefficients satisfy conjugate symmetry. It is therefore sufficient to estimate the coefficients at the nonnegative frequencies $k=0,\ldots,L$.
To break the inherent rotational symmetry of the MRA problem, we arbitrarily fix the phase of the $k=1$ coefficient by setting~$\widehat{\hf}[1] = \hat{r}_1$.

\subsubsection{Low-SNR expansion of the likelihood function}
\label{subsec:conditional_likelihood}
To estimate the Fourier phases, we optimize the marginalized log-likelihood with respect to one phase variable at a time, while holding the remaining coefficients fixed. Our starting point is the full marginalized log-likelihood: for a candidate Fourier vector $\hg=(\hg[0],\ldots,\hg[L])$, we define the log-likelihood, up to constants independent of $\hg$, by
\begin{equation}
	\label{eq:L_so2_2}
	\mathcal{L}(\hg) = \frac{1}{n} \sum_{j=1}^n \log\int_0^{2\pi} e^{ -\frac{1}{2\sigma^2} \sum_{m=0}^L \left| \hy_j[m]-\hg[m]e^{-im\theta}\right|^2 } d\theta.
\end{equation}
We then define a family of phase-restricted likelihoods, one for each frequency $k$. To estimate the phase of the $k$-th Fourier coefficient, we fix the current values of all other Fourier coefficients, as well as the magnitude of the $k$-th coefficient, and optimize only over its phase. For $|\phi|=1$, we define
\begin{equation}
	\mathcal{L}_k(\phi) = \mathcal{L}\left( \hf[0], \ldots, \hf[k-1], \, r_k \phi \, , \hf[k+1], \ldots, \hf[L] \right).
\end{equation}
The corresponding phase update is the restricted maximum-likelihood problem    
\begin{equation}
	\widehat\phi[k]\in\argmax_{|\phi|=1}\mathcal{L}_{k}(\phi).
\end{equation}

Expanding the squared norm in \eqref{eq:L_so2_2}, we have
\begin{equation}
	\label{eqn:log-likelhood-simplified}
	\mathcal{L}_{k}(\phi)
	= C_{k} + \frac{1}{n} \sum_{j=1}^n \log\int_0^{2\pi} H_{j,k}(\theta) e^{ \frac{|\hf[k]|}{\sigma^2} \Re\left( \overline{\hy_j[k]} \,\phi e^{-ik\theta} \right)} d\theta,
\end{equation}
where $C_k$ is independent of $\phi$, and
\begin{equation}
	\label{eq:H_jk}
	H_{j,k}(\theta) := e^{\frac{1}{\sigma^2} \sum_{k'\neq k} \Re\left( \overline{\hy_j[k']} \,\hf[k']e^{-ik'\theta} \right) }.
\end{equation}

The following lemma gives the low-SNR expansion of the restricted likelihood~\eqref{eqn:log-likelhood-simplified}. Its proof is given in Appendix~\ref{app:lemma_L_so2}. The expansion is taken with respect to the contribution of the current phase variable, namely the second exponential in~\eqref{eqn:log-likelhood-simplified}, while all other coefficients are kept fixed. Consequently, the phase-dependent leading term is linear in $\phi$, and the phase update has a closed-form solution.

\begin{lemma}[Low-SNR restricted log-likelihood over~$\SO(2)$]
	\label{lem:L_so2}
	Fix $k\in \{1,\ldots,L\}$, and hold $r_k$ and the coefficients $\{\hf[k']: \,k'\neq k\}$ fixed. Define
	\begin{equation}
		s_{j,k}[m] := \int_{0}^{2\pi} H_{j,k}(\theta) e^{-im\theta} d\theta.
		\label{eq:sjk}
	\end{equation}
	Then $s_{j,k}[0]$ is real, and strictly positive, and the restricted log-likelihood~\eqref{eqn:log-likelhood-simplified} satisfies
	\begin{equation}
		\mathcal{L}_{k}(\phi) = \tilde{C}_k + \frac{r_k}{\sigma^2} \Re\left\{ \overline{Z_k} \, \phi \right\} + R_{k}(\phi),        \label{eq:L_so2_low_snr}
	\end{equation}
	where $\tilde{C}_k$ is independent of $\phi$,
	\begin{equation}
		Z_k := \frac{1}{n} \sum_{j=1}^n \hy_j[k] \, \frac{\overline{s_{j,k}[k]}}{s_{j,k}[0]},
		\label{eq:Zk_so2}
	\end{equation}
	and, uniformly over $|\phi|=1$, $R_{k}(\phi) = \mathcal{O}_{\mathbb{P}}(r_k^2/\sigma^2)$.\footnote{For a stochastic sequence $X_n$ and a deterministic positive sequence~$a_n$, the notation~$X_n=\mathcal{O}_{\mathbb{P}}(a_n)$ means that~$X_n/a_n$ is bounded in probability as~$n\to\infty$.}
\end{lemma}

In particular, the phase-dependent leading term of the restricted log-likelihood is~$\frac{r_k}{\sigma^2} \Re\left\{ \overline{Z_k} \, \phi \right\}$.
Therefore, up to the higher-order remainder in the low-SNR expansion, the phase update is equivalent to
\begin{equation}
	\widehat{\phi}[k] \in \argmax_{|\phi|=1} \Re\left\{  \overline{Z_k}\, \phi \right\}. \label{eq:phase_update_optimization}
\end{equation}
Whenever $Z_k\neq 0$, this maximization has the simple closed-form solution
\begin{equation}
	\widehat{\phi}[k] = Z_k/|Z_k|. \label{eq:solution_phase}
\end{equation}

\subsubsection{The algorithm}
The fast approximate MLE algorithm is summarized in Algorithm~\ref{alg:fast_mle}.
First, we estimate the magnitudes of all~$L$ coefficients and the zero-frequency coefficient using~\eqref{eq:magnitudes} and~\eqref{eq:avg}, respectively.
In the frequency-marching implementation, we initialize the coefficients at frequencies $k\geq 2$ to zero and then recover their phases sequentially from lower to higher frequencies.

Although the magnitudes of the higher-frequency coefficients have already been estimated, setting them to zero during the sweep suffices for consistent recovery: the leading-order update at frequency~$k$ couples the next component only to previously recovered lower components~\eqref{eq:solution_phase}, so each phase is identified sequentially (Section~\ref{sec:consistency_so2}). These higher frequencies may still carry finite-sample information and can be incorporated through additional refinement sweeps.

\begin{algorithm}[tb]
	\caption{Fast Approximate MLE Algorithm}
	\label{alg:fast_mle}
	\begin{algorithmic}[1]
		
		\REQUIRE Observations $\{\hy_j\}_{j=1}^n$, noise variance $\sigma^2$, bandlimit $L$.
		\ENSURE Estimated Fourier coefficients $\widehat \hf$.
		
		\textbf{Initialize:}
		\STATE Estimate the magnitudes~$\{\hat{r}_k\}_{k=1}^L$ according to~\eqref{eq:magnitudes} and~$\widehat\hf[0]$ according to~\eqref{eq:avg}.
		\STATE $\widehat \hf[1] \leftarrow \hat{r}_1$
		\STATE $\widehat \hf[k] \leftarrow 0, \quad \forall k \in \{2, \ldots ,L\}$
		\FOR{$k = 2$ \TO $L$}
		\FOR{$j = 1$ \TO $n$}
		\STATE \textit{Compute  $s_{j,k}$ according to~\eqref{eq:sjk}}
		\ENDFOR
		\STATE \textit{Compute optimal phase according to~\eqref{eq:solution_phase} and update the~$k$-th coefficient:}
		\STATE $\widehat\phi[k] \leftarrow Z_k / |Z_k|$, $\widehat \hf[k] \leftarrow \hat{r}_k \widehat \phi[k]$
		\ENDFOR
		\RETURN $\widehat \hf$
	\end{algorithmic}
\end{algorithm}

\begin{remark}[A single-pass variant]
	\label{rem:single_pass_variant}
	A direct, single-pass variant is obtained by building the restricted likelihood from only the fixed first Fourier coefficient: for each~$k\geq2$, replace the kernel~$H_{j,k}$ in~\eqref{eq:H_jk} by $H_j^{(1)}(\theta):=e^{\frac{1}{\sigma^2}\Re(\overline{\hy_j[1]}\hf[1]e^{-i\theta})}$ and apply the same first-order expansion as in Lemma~\ref{lem:L_so2}. This estimates all phases in a single pass over the data. In our experiments it was less accurate than the frequency-marching update of Algorithm~\ref{alg:fast_mle}, which attains higher accuracy at negligible additional cost on a given dataset; moreover, it has weaker sample-complexity guarantees than the marching algorithm (Remark~\ref{rem:single_pass_rate}); we therefore adopt the marching version.
\end{remark}

\subsection{Consistency of the frequency marching estimator}
\label{sec:consistency_so2}
We state the consistency of the proposed frequency marching estimator in the low-SNR asymptotic regime.

\begin{thm}[Consistency over~$\SO(2)$]
	\label{thm:consistency_so2}
	Consider the~$\SO(2)$ MRA model~\eqref{eq:observations} with fixed bandlimit~$L$ and fixed signal~$\hf$ satisfying~$\hf[k]\neq0$ for all~$k=1,\ldots,L$. Suppose that~$n=\omega(\sigma^6)$ as~$\sigma\to\infty$. Then, the frequency marching estimator of Algorithm~\ref{alg:fast_mle} satisfies~$\widehat{\hf}\xrightarrow{p}\hf$, up to a global rotation.\footnote{The notation~$\widehat{\hf}\xrightarrow{p}\hf$ denotes convergence in probability: for every~$\epsilon>0$, $P\{\|\widehat{\hf}-\hf\|_2>\epsilon\}\to0$.}
\end{thm}

The proof is given in Appendix~\ref{app:proof_consistency_so2}. The argument proceeds by induction over the frequency index~$k$. The preliminary estimates of~$\hf[0]$ and of the magnitudes~$r_k$ are consistent under the weaker conditions~$n=\omega(\sigma^2)$ and~$n=\omega(\sigma^4)$, respectively. It remains to prove consistency of the phases.

Assume that the lower-frequency coefficients have already been consistently estimated. The low-SNR expansion of the~\mbox{$k$-th} likelihood update shows that its leading term is a weighted third-order statistic of the form~$\hy_j[k]\overline{\hy_j[a]}\overline{\hy_j[b]}$, where~$a+b=k$. The rotation phases cancel, so the population update is a positive real multiple of~$\hf[k]$. The leading expectation is of order~$\sigma^{-4}$, while the fluctuation is of order~$1/(\sigma\sqrt n)$. Hence, after multiplication by~$\sigma^4$, the fluctuation vanishes precisely when~$n=\omega(\sigma^6)$. Therefore, the estimated phase converges to the phase of~$\hf[k]$, and the result follows by induction.

This calculation also clarifies the relation to bispectrum inversion~\cite{bendory2017bispectrum}. In the strict low-SNR limit, the leading term of the likelihood-based update contains the same phase information as the bispectrum entries~$B_f[a,b]=\hf[a]\hf[b]\overline{\hf[k]}$, where~$a+b=k$. Thus, the proposed update recovers the same leading third-order invariant information as the bispectrum, but it is derived directly from the expansion of the marginalized likelihood.

\subsection{Computational complexity analysis for MRA over~$\SO(2)$}
\label{sec:complexity}

In the following, we compare the computational complexity of the proposed fast approximate MLE with EM and bispectrum inversion in the~$\SO(2)$ setting. Let~$n$ be the number of measurements and let~$L$ be the bandlimit. The preliminary magnitude estimation requires averaging over the measurements for each frequency, and therefore costs~$\mathcal{O}(nL)$ operations. The proposed algorithm then estimates the Fourier phases sequentially. For each frequency~$k\in\{2,\ldots,L\}$, the quantities~$s_{j,k}[m]$ in~\eqref{eq:sjk} are evaluated by numerical integration over a grid of rotation angles in~$[0,2\pi)$. We denote by~$N_{\theta}^{(\mathrm{fast})}$ the number of grid points used for this numerical integration.
In the implementation of Algorithm~\ref{alg:fast_mle}, the unresolved higher-frequency coefficients are set to zero. Therefore, when estimating the phase of the $k$-th coefficient, the sum inside the exponent in~\eqref{eq:sjk} involves only the previously recovered lower-frequency coefficients. This exponent can be updated incrementally on the angular grid: when moving from frequency~$k$ to frequency~$k+1$, one adds the contribution of the newly recovered $k$-th coefficient, rather than recomputing the entire sum from scratch. Thus, one forward sweep of the proposed method has dominant cost~$\mathcal{O}(nL N_{\theta}^{(\mathrm{fast})})$.

For comparison, EM approximates the marginalization over~$\SO(2)$ using a rotation grid at every iteration~\cite[Section VI]{bendory2017bispectrum}. If~$N_{\theta}^{(\mathrm{EM})}$ denotes the number of grid points used in the EM E-step and~$T_{\mathrm{EM}}$ denotes the number of EM iterations, then the total complexity of EM is~$\mathcal{O}(T_{\mathrm{EM}}nL N_{\theta}^{(\mathrm{EM})})$. Crucially,~$T_{\mathrm{EM}}$ is not constant and  grows rapidly as the SNR decreases~\cite{janco2022accelerated,balanov2025expectation}. The bispectrum-based invariant method is the most efficient algorithm in the~$\SO(2)$ setting, with coefficient-domain complexity~$\mathcal{O}(nL)$~\cite{bendory2017bispectrum}. This efficiency is possible because the method avoids both iterative refinement and numerical integration over the rotation group. As we shall see, the corresponding invariant approach becomes substantially more expensive in the~$\SO(3)$ setting, where the bispectrum involves coupled spherical-harmonic degrees and radial indices.

\section{Fast MLE algorithm for low-SNR MRA over~$\SO(3)$}
\label{sec:so3}
In this section, we extend the likelihood-expansion framework from~$\SO(2)$ to~$\SO(3)$, the focus of this work. We show that the same low-SNR approximation used in the~$\SO(2)$ case leads to a closed-form Procrustes update. We then describe EM and a bispectrum-based invariant method as baselines, and compare the computational complexity of the three approaches. Section~\ref{sec:numerical_experiments} presents a numerical comparison of these methods on synthetic $\SO(3)$ MRA observations generated from molecular volumes, evaluating both reconstruction accuracy and running time.

\subsection{Statistical model}
We consider a real function~$X$ on the ball~$B_{R_\mathrm{ball}}=\{x\in\R^3: \|x\|\leq R_\mathrm{ball}\}$. Let~$\{\psi_{\ell,q}\}_{q=1}^{R}$ be a family of radial basis functions associated with angular degree~$\ell$, where~$R$ denotes the number of radial degrees of freedom per angular degree, and let~$L$ be the angular bandlimit. We represent~$X$ by the finite-dimensional radial spherical-harmonic expansion
\begin{equation} 
	\label{eq:signal_so3}
	X(r,\theta,\varphi) = \sum_{\ell=0}^{L} \sum_{m=-\ell}^{\ell} \sum_{q=1}^{R} F_{\ell,m,q} \, \psi_{\ell, q}(r) \, \mathcal{Y}_{\ell}^{m}(\theta,\varphi),
\end{equation}	where~$\mathcal{Y}_\ell^m$ are the complex spherical harmonic basis functions.

For each degree~$\ell = 0,\ldots,L$, we collect the coefficients into the matrix
\begin{equation}
	\label{eq:Fell_definition}
	F_\ell := \left(F_{\ell,m,q}\right)_{\substack{m=-\ell,\ldots,\ell \\ q=1,\ldots,R}} \in \mathbb{C}^{(2\ell+1)\times R},
\end{equation}
whose rows are indexed by the spherical-harmonic order~$m$ and whose columns are indexed by the radial-basis index~$q$.    
The collection~$\{F_\ell\}_{\ell=0}^{L}$ is equivalent to the truncated coefficient-domain representation in~\eqref{eq:signal_so3}.
Thus, after fixing the radial and angular bases, estimating the truncated volume is equivalent to estimating the coefficient matrices~$\{F_\ell\}_{\ell=0}^{L}$ from the noisy rotated observations. We assume~$R\geq 2L+1$, so that, for every~$\ell\leq L$, the matrix~$F_\ell\in\mathbb{C}^{(2\ell+1)\times R}$ is generically full row rank~\cite{bendory2025orbit}. This ensures that the second moment~$F_\ell^{\ast}F_\ell$ determines~$F_\ell$ up to a left-acting unitary matrix, which is the parametrization used below. 
The radial spherical-harmonic expansion in~\eqref{eq:signal_so3} allows for general radial bases, including the spherical-Bessel bases used in the numerical experiments.
Since the volume is real-valued, the complex spherical-harmonic coefficients satisfy the conjugate-symmetry relation $F_{\ell,-m,q} = (-1)^m \overline{F_{\ell,m,q}}$.

Rotating~$X$ by~$\omega\in\SO(3)$ acts by
\begin{equation}
	\label{eq:rotation}
	(R_\omega X)_{\ell,m,q} =
	\sum_{m'=-\ell}^{\ell} D_{m,m'}^\ell(\omega)\, F_{\ell,m',q},
\end{equation}    
where~$D_{m,m'}^\ell(\omega)$ are the entries of the Wigner-$D$ matrix of degree~$\ell$. Defining~$D_\ell(\omega)\in\mathbb{C}^{(2\ell+1)\times(2\ell+1)}$, the rotation action on the degree-$\ell$ coefficient matrix is expressed compactly as $F_\ell \mapsto D_\ell(\omega)F_\ell$.

We consider a set of $n$ measurements
\begin{equation}
	\label{eq:measurements}
	y_j = R_{\omega_j}X+\epsilon_j,
	\qquad j=1,\ldots,n,
\end{equation}
where $\omega_j\stackrel{\mathrm{i.i.d.}}{\sim}\mathrm{Haar}(\SO(3))$.
In the finite-dimensional coefficient representation defined in~\eqref{eq:signal_so3}, the observations satisfy
\begin{equation}
	\label{eq:measurements_so3}
	Y_{j,\ell} = D_\ell(\omega_j)F_\ell+\mathcal{E}_{j,\ell},
\end{equation}
where $Y_{j,\ell}, \mathcal{E}_{j,\ell}\in\mathbb{C}^{(2\ell+1)\times R}$, and the entries of $\mathcal{E}_{j, \ell}$ are independent complex Gaussian random variables with mean zero and variance $\sigma^2$, subject to the conjugate-symmetry constraints induced by the real-valuedness of the volume.
As in the~$\SO(2)$ case, the signal can be estimated only up to a global~$\SO(3)$ rotation.

\subsection{Fast approximate MLE for low SNR MRA over~$\SO(3)$}
\label{sec:MLE_so3}

The proposed estimator follows the same high-level strategy as in the $\SO(2)$ setting. We first use rotation-invariant first and second-order moments to estimate the degree-zero coefficient and the Gram matrix of each spherical-harmonic band. These estimates determine each coefficient matrix up to a left-acting unitary transformation. We then recover the remaining unitary matrices sequentially from lower to higher degrees. The  algorithm is summarized in Algorithm~\ref{alg:fast_mle_so3}.
\subsubsection{Preliminary estimates}
The degree-zero coefficients are invariant to rotations, since~$D_0(\omega)=1$. We therefore estimate~$F_0$ by averaging, which is consistent when~$n=\omega(\sigma^2)$:
\begin{equation}
	\label{eq:F0_so3}
	\widehat{F}_0 = \frac{1}{n}\sum_{j=1}^n Y_{j,0}.
\end{equation}

Next, we introduce a factorization of the coefficients matrices~$\{F_\ell\}_{\ell=0}^L$. The second moment matrices
\begin{equation}
	M_\ell := F_\ell^{\ast} F_\ell \in \mathbb{C}^{R \times R}
\end{equation} 
are rotation-invariant. Since $\mathbb{E}[Y_{j,\ell}^{\ast} Y_{j,\ell}] = M_\ell + (2\ell+1)\sigma^2 I_R$, we estimate
\begin{equation} 
	\label{eq:second_moment}
	\widehat{M}_\ell = \frac{1}{n} \sum_{j=1}^n Y_{j,\ell}^{\ast} Y_{j,\ell} - (2\ell+1)\sigma^2 I_R.
\end{equation}

Since~$F_\ell \in \mathbb{C}^{(2\ell+1) \times R}$ with~$R \geq 2\ell+1$, the matrix~$M_\ell = F_\ell^{\ast} F_\ell \in \mathbb{C}^{R \times R}$ has rank at most~$2\ell+1$ and determines~$F_\ell$ up to a left-acting unitary matrix~$Q_\ell \in \U(2\ell+1)$. We construct~$\tilde{F}_\ell$ from $\widehat{M}_\ell$ via eigendecomposition. Let $\lambda_1 \geq \cdots \geq \lambda_R$ be the real eigenvalues of~$\widehat{M}_\ell$, with corresponding eigenvectors collected in~$V\in\mathbb{C}^{R\times R}$. We retain the leading~$2\ell+1$ eigenvectors, denoted by~$V_{2\ell+1}\in\mathbb{C}^{R\times(2\ell+1)}$, and truncate possible negative eigenvalues.
We set
\begin{equation}
	\label{eq:tilde_F}
	\widetilde{F}_\ell = \Sigma_\ell^{1/2} V_{2\ell+1}^{\ast},	\Sigma_\ell := \mathrm{diag} \left(\lambda_1^{+},\ldots,\lambda_{2\ell+1}^{+}\right),
\end{equation}
where $x^{+}\triangleq\max\{x,0\}$.

At the population level, let ${\Fpop}_\ell$ denote the counterpart of the empirical factor $\widetilde{F}_\ell$, that is $\widetilde{F}_\ell \xrightarrow[]{\mathbb{P}} {\Fpop}_\ell$, as $n \to\infty$. Thus, by the construction in~\eqref{eq:tilde_F}, ${\Fpop}_\ell^\ast {\Fpop}_\ell = M_\ell$. Since $F_\ell^\ast F_\ell = M_\ell$, the matrices $F_\ell$ and ${\Fpop}_\ell$ have the same Gram matrix. Therefore, there exists a unitary matrix $Q_\ell \in \mathrm{U}(2\ell+1)$ such that
\begin{equation}
	F_\ell = Q_\ell {\Fpop}_\ell. \label{eq:decomposition}
\end{equation}
Thus, after estimating the Gram matrices, the remaining task is to recover the unitary matrices
$\{Q_\ell\}_{\ell=0}^{L}$. These matrices play the same role as the unresolved Fourier phases in the $\SO(2)$ setting.

The degree-zero component is estimated directly by averaging, as in~\eqref{eq:F0_so3}. For the remaining degrees, the coefficient-domain observation model~\eqref{eq:measurements_so3} can be written as
\begin{equation}
	Y_{j,\ell} = D_\ell(\omega_j) Q_\ell {\Fpop}_\ell + \mathcal{E}_{j,\ell},
	\qquad
	j = 1,\ldots,n. \label{eq:measurements_normalized}
\end{equation}
In practice, ${\Fpop}_\ell$ in~\eqref{eq:measurements_normalized} is replaced by its estimate $\widetilde{F}_\ell$.

Since the $\ell=0$ Wigner-D matrix satisfies $D_0(\omega)=1$ for all $\omega\in\SO(3)$, the $\ell=0$ measurements reduce to $Y_{j,0} = F_0 + \mathcal{E}_{j,0}$, independent of the unknown rotations. Thus $Q_0 \in \{+1,-1\}$ is a sign, determined by comparing the mean estimator~\eqref{eq:F0_so3} with $\tilde{F}_0$ obtained from~\eqref{eq:tilde_F}:
\begin{equation}
	\label{eq:Q0}
	\widehat{Q}_0 = \operatorname{sign}\left(\Re \widehat{F}_0 \overline{\widetilde{F}_0} \right).
\end{equation}
This resolves the sign ambiguity introduced by the eigendecomposition in~\eqref{eq:tilde_F}.

To break the inherent rotational symmetry of the problem, we fix the~$Q_1$ matrix, while preserving the conjugate symmetry property. The canonical unitary choice is
\begin{equation} 
	\label{eq:Q1}
	\widehat{Q}_1 = \begin{pmatrix} -\tfrac{1}{\sqrt{2}} & 0 & \tfrac{i}{\sqrt{2}} \\[3pt] 0 & 1 & 0 \\[3pt] \tfrac{1}{\sqrt{2}} & 0 & \tfrac{i}{\sqrt{2}} \end{pmatrix}.
\end{equation}

\begin{remark}[$O(3)$ component ambiguity]    \label{rem:o3_component_ambiguity}
	The second moment determines the degree-one coefficient matrix~$F_1$ only up to an element of~$O(3)$, leaving two possible~$\SO(3)$ orbits, corresponding to the orientation-preserving and orientation-reversing components.
	Rather than selecting the correct component using a separate invariant test, as done for example in~\cite{bendory2025orbit}, we run the subsequent marching procedure for both representatives and retain the reconstruction with the larger marginalized likelihood.
\end{remark}

\subsubsection{Low-SNR expansion of the likelihood function}

As in the~$\SO(2)$ case, we optimize the marginalized log-likelihood with respect to one unresolved variable at a time, while holding the remaining coefficient matrices fixed.
Here the scalar phase variable is replaced by the unitary matrix~$Q_\ell$.

For a candidate collection of coefficient matrices~$G=\{G_\ell\}_{\ell=0}^{L}$, we define the log-likelihood, up to constants independent of~$G$, by
\begin{equation}
	\label{eq:L_so3_full}
	\mathcal{L}(G) = \frac{1}{n} \sum_{j=1}^{n} \log \int_{\SO(3)} e^{ -\frac{1}{2\sigma^2} \sum_{\ell=0}^{L} \left\| Y_{j,\ell}-D_\ell(\omega)G_\ell \right\|_{\F}^{2} } d\omega,
\end{equation}
where~$d\omega$ denotes the normalized Haar measure on~$\SO(3)$ and~$\| \cdot \|_{\text{F}}$ is the Frobenius norm.

We then define a family of restricted likelihoods, one for each degree~$\ell$.
To estimate~$Q_\ell$, we fix~$\widetilde{F}_\ell$ and the current values of all other coefficient matrices, and optimize only over~$Q_\ell$.
Specifically, for~$Q\in\U(2\ell+1)$, we define
\begin{equation}
	\mathcal{L}_\ell(Q) = \mathcal{L} \left( F_0, \ldots,  F_{\ell-1}, Q\widetilde{F}_\ell, F_{\ell+1}, \ldots, F_L \right).
\end{equation}
The corresponding restricted maximum-likelihood problem is
\begin{equation}
	\widehat{Q}_\ell
	\in
	\argmax_{Q\in\U(2\ell+1)}
	\mathcal{L}_\ell(Q).
\end{equation}

Expanding the squared Frobenius norms in~\eqref{eq:L_so3_full}, all terms that do not involve~$Q$ can be absorbed into a constant independent of~$Q$.
Thus,
\begin{equation}
	\label{eq:L_so3_restricted}
	\mathcal{L}_\ell(Q) = C_\ell + \frac{1}{n} \sum_{j=1}^{n} \log \int_{\SO(3)}  H_{j,\ell}(\omega)  e^{ \frac{1}{\sigma^2} \Re\tr\left( \widetilde{F}_\ell Y_{j,\ell}^{*}D_\ell(\omega)Q \right) } d\omega,
\end{equation}
where~$C_\ell$ is independent of~$Q$, and
\begin{equation}
	H_{j,\ell}(\omega) := e^{\frac{1}{\sigma^2} \sum_{\ell'\neq \ell}\Re\tr\left( F_{\ell'}Y_{j,\ell'}^{*}D_{\ell'}(\omega) \right) }.        \label{eq:so3_alignment_kernel}
\end{equation}
The next lemma gives the low-SNR approximation of the restricted likelihood~\eqref{eq:L_so3_restricted}.
Its proof is given in Appendix~\ref{app:lemma_L_so3}.

\begin{lemma}[Low-SNR restricted log-likelihood over~$\SO(3)$]
	\label{lem:L_so3}
	Fix~$\ell\in \{0,\ldots,L\}$, and hold~$\widetilde{F}_\ell$ and the current coefficient matrices~$\{F_{\ell'}:\ell'\neq\ell\}$ fixed.
	For each representation degree~$m$, define
	\begin{equation}
		s_{j,\ell}[m] := \int_{\SO(3)} H_{j,\ell}(\omega)D_m(\omega)\,d\omega .
		\label{eq:s_so3}
	\end{equation}
	Since~$D_0(\omega)=1$, the quantity $s_{j,\ell}[0] = \int_{\SO(3)} H_{j,\ell}(\omega) \,d\omega$ is real and strictly positive, and the restricted log-likelihood in~\eqref{eq:L_so3_restricted} satisfies
	\begin{equation}
		\mathcal{L}_\ell(Q) = \tilde{C}_\ell + \frac{1}{\sigma^2} \Re\tr\left( A_\ell Q \right) + R_\ell(Q),
		\label{eq:L_so3_low_snr}
	\end{equation}
	where~$\tilde{C}_\ell$ is independent of~$Q$,
	\begin{equation}
		A_\ell := \frac{1}{n} \widetilde{F}_\ell \sum_{j=1}^{n} \frac{Y_{j,\ell}^{*}s_{j,\ell}[\ell]}{s_{j,\ell}[0]} ,
		\label{eq:Aell_so3}
	\end{equation}
	and, uniformly over~$Q\in\U(2\ell+1)$, $R_\ell(Q) = \mathcal{O}_{\mathbb{P}} (\|\widetilde{F}_\ell\|_{\F}^{2}/\sigma^2)$.
\end{lemma}

In particular, the leading $Q$-dependent term of the restricted log-likelihood
is~$\frac{1}{\sigma^2}\Re\tr(A_\ell Q)$. The matrix~$A_\ell$~\eqref{eq:Aell_so3} carries the known left factor~$\widetilde{F}_\ell$; dividing it out gives the normalized matrix
\begin{equation}
	\bar{A}_\ell := \Sigma_\ell^{\dagger} A_\ell,
	\label{eq:Aell_whitened_so3}
\end{equation}
where~$\Sigma_\ell = \widetilde{F}_\ell\widetilde{F}_\ell^{*}$ is already available from~\eqref{eq:tilde_F} and~$\dagger$ denotes the Moore--Penrose pseudoinverse. To achieve consistency of the resulting estimator (Section~\ref{sec:consistency_so3}), we introduce this whitening factor: the population limit of~$A_\ell$ retains~$\breve{\Sigma}_\ell$ (see~\eqref{eq:Aell_pop_unwhitened} in Appendix~\ref{app:proof_consistency_so3}), and left-multiplication by~$\Sigma_\ell^{\dagger}$ removes it. The update for~$Q_\ell$ is then the unitary Procrustes problem
\begin{equation}
	\widehat{Q}_\ell \in \argmax_{Q\in\U(2\ell+1)} \Re\tr\!\left(\bar{A}_\ell\, Q\right),
	\label{eq:procrustes_so3}
\end{equation}
with solution
\begin{equation}
	\widehat{Q}_\ell = V_\ell U_\ell^{*}
	\label{eq:solution_Q_so3}
\end{equation}
from the SVD~$\bar{A}_\ell = U_\ell S_\ell V_\ell^{*}$. Finally, the degree-$\ell$ coefficient matrix is estimated by
\begin{equation}
	\widehat{F}_\ell =  \widehat{Q}_\ell\widetilde{F}_\ell.
\end{equation}

\subsubsection{The algorithm}

\begin{algorithm}[tb]
	\caption{Fast Approximate MLE Algorithm over $\SO(3)$}
	\label{alg:fast_mle_so3}
	\begin{algorithmic}[1]
		
		\REQUIRE Observations $\{Y_{j}\}_{j=1}^{n}$, noise variance $\sigma^2$, bandlimit $L$.
		\ENSURE Estimated coefficient matrices $\{\widehat{F}_\ell\}_{\ell=0}^L$.
		
		\textbf{Initialize:}
		\STATE Estimate~$\widehat{F}_0$ according to~\eqref{eq:F0_so3}.
		\STATE Estimate~$\widehat{M}_\ell$ according to~\eqref{eq:second_moment}
		and compute~$\widetilde{F}_\ell$ and~$\Sigma_\ell$ according to~\eqref{eq:tilde_F}.
		\STATE Initialize~$\widehat{Q}_\ell\leftarrow 0$ for all~$\ell=0,\ldots,L$.
		\STATE Set~$\widehat{Q}_0$ according to~\eqref{eq:Q0}
		and fix~$\widehat{Q}_1$ according to~\eqref{eq:Q1} to remove the
		global rotational ambiguity.
		\STATE Set~$\widehat{F}_\ell\leftarrow \widehat{Q}_\ell\widetilde{F}_\ell$
		for~$\ell=0,\ldots,L$.
		\FOR{$\ell=2$ \TO $L$}
		\FOR{$j=1$ \TO $n$}
		\STATE Estimate~$s_{j,\ell}[0]$ and~$s_{j,\ell}[\ell]$~\eqref{eq:s_so3}
		by Monte Carlo integration over~$\SO(3)$ using the current
		estimates~$\{\widehat{Q}_q:q\neq\ell\}$.
		\ENDFOR
		\STATE Compute~$\bar{A}_\ell \leftarrow \Sigma_\ell^{\dagger}A_\ell$ according to~\eqref{eq:Aell_whitened_so3}.
		\STATE Compute an SVD~$\bar{A}_\ell = U_\ell S_\ell V_\ell^{\ast}$.
		\STATE Compute $\widehat{Q}_\ell\leftarrow V_\ell U_\ell^{\ast}$, and update~$\widehat{F}_\ell\leftarrow \widehat{Q}_\ell\widetilde{F}_\ell$.
		\ENDFOR
		
		\RETURN $\{\widehat{F}_\ell\}_{\ell=0}^L$.
		
	\end{algorithmic}
\end{algorithm}

The fast approximate MLE algorithm over~$\SO(3)$ is summarized in Algorithm~\ref{alg:fast_mle_so3}. First, we estimate the degree-zero coefficients~$F_0$ and the second-moment factors~$\widetilde{F}_\ell$ using~\eqref{eq:F0_so3} and~\eqref{eq:tilde_F}, respectively. We then set the fixed gauge bands:~$Q_0$ is determined from the degree-zero estimate, and~$Q_1$ is fixed to remove the global rotational ambiguity. Similar to the~$\SO(2)$ case, the algorithm proceeds by frequency marching, sweeping over the spherical-harmonic degrees~$\ell=2,\ldots,L$. During the first forward sweep, unresolved higher-degree coefficient matrices are set to zero, even though their Gram factors~$\widetilde{F}_q$ have already been estimated. This avoids introducing arbitrary unitary matrices~$Q_q$ before they have been inferred. Thus, when updating degree~$\ell$, the alignment kernel~$H_{j,\ell}$~\eqref{eq:so3_alignment_kernel} is constructed using the already recovered lower-degree coefficient matrices, while the unresolved higher degrees remain inactive. The quantities~$s_{j,\ell}[0]$ and~$s_{j,\ell}[\ell]$~\eqref{eq:s_so3} are then estimated by Monte Carlo integration over~$\SO(3)$, the normalized matrix~$\bar{A}_\ell$ in~\eqref{eq:Aell_whitened_so3} is formed, and~$\widehat{Q}_\ell$ is updated by the Procrustes solution~\eqref{eq:solution_Q_so3}. Finally, the degree-$\ell$ coefficient matrix is updated as~$\widehat{F}_\ell=\widehat{Q}_\ell\widetilde{F}_\ell$. As in the~$\SO(2)$ case, omitting the unresolved higher degrees during the forward sweep does not prevent recovery in the low-SNR marching regime, and they can be incorporated through additional refinement sweeps if desired.

\subsection{Consistency of the frequency marching estimator over~$\SO(3)$}
\label{sec:consistency_so3}
We next state the analogous consistency result for the proposed estimator over~$\SO(3)$. Since the signal is identifiable only up to a global rotation, we work in the fixed global gauge determined by the prescribed choice of~$Q_1$, with the $O(3)$ component ambiguity handled as described in Remark~\ref{rem:o3_component_ambiguity}.

\begin{thm}[Consistency over~$\SO(3)$]
	\label{thm:consistency_so3}
	Consider the~$\SO(3)$ MRA model~\eqref{eq:measurements_so3} with fixed angular bandlimit~$L$ and fixed radial dimension~$R\geq 2L+1$. Suppose that the coefficient matrices~$\{F_\ell\}_{\ell=0}^L$ are generic and full row rank. Suppose that~$n=\omega(\sigma^6)$ as~$\sigma\to\infty$. Then, the frequency marching estimator of Algorithm~\ref{alg:fast_mle_so3} satisfies~$\widehat{F}_\ell\xrightarrow{p}F_\ell$,~$\ell=0,\ldots,L$, up to a global rotation.
\end{thm}

The proof is given in Appendix~\ref{app:proof_consistency_so3}. The argument parallels the~$\SO(2)$ proof. This proof also clarifies the relation to the~$\SO(3)$ bispectrum (Section~\ref{sec:bispectrum_so3}). In the low-SNR limit, the leading likelihood-based update depends on the same third-order rotational invariants as the bispectrum.

\subsection{Expectation-Maximization (EM) for MRA over~$\SO(3)$}
\label{sec:em_so3}
EM is the standard likelihood-based iterative approach for approximating the MLE in latent-variable models such as MRA~\cite{sigworth1998maximum}. In the present setting, the latent variables are the unknown 3D rotations. EM alternates between estimating their posterior distribution under the current signal estimate and updating the signal by maximizing the corresponding expected log-likelihood. To approximate the marginalization over~$\SO(3)$, we discretize the rotation group. Let~$\Omega^{(\mathrm{EM})}$ be a sampled set of rotations in~$\SO(3)$. Let~$\{\widehat{F}_\ell^{(t)}\}_{\ell=0}^{L}$ denote the signal estimate at iteration~$t$.   In the E-step, EM computes the posterior weights of the rotations over the discrete grid. For the~$j$-th observation and grid point~$\omega_k$, the weight is
\begin{equation}
	\label{eq:e_step_so3}
	\gamma_j^{(t)}(\omega_k) = \frac{ e^{ -\frac{1}{2\sigma^2} \sum_{\ell=0}^{L} \left\| Y_{j,\ell}-D_\ell(\omega_k)\widehat{F}_\ell^{(t)} \right\|_{\F}^{2} } }{ \sum_{\omega_{k'} \in \Omega^{(\mathrm{EM})}} e^{ -\frac{1}{2\sigma^2} \sum_{\ell=0}^{L} \left\| Y_{j,\ell}-D_\ell(\omega_{k'})\widehat{F}_\ell^{(t)} \right\|_{\F}^{2} } },
\end{equation}
so that~$\sum_{\omega_{k} \in \Omega^{(\mathrm{EM})}}\gamma_j^{(t)}(\omega_k)=1$. In the M-step, the coefficient matrices are updated by maximizing the expected complete-data log-likelihood.
This yields the weighted average of the inverse-rotated observations
\begin{equation}
	\label{eq:m_step_so3} \widehat{F}_\ell^{(t+1)} = \frac{1}{n} \sum_{j=1}^{n} \sum_{\omega_{k} \in \Omega^{(\mathrm{EM})}} \gamma_j^{(t)}(\omega_k) D_\ell(\omega_k)^{*}Y_{j,\ell}.
\end{equation}

\subsection{Bispectrum-based invariant method for MRA over~$\SO(3)$}
\label{sec:bispectrum_so3}
We use the bispectrum-based invariant baseline of~\cite{bendory2025orbit}, which estimates rotation-invariant quantities directly from the observations by coupling spherical-harmonic coefficients of different degrees through Clebsch--Gordan coefficients, without estimating the latent rotations. Recall that~$Y_{j,\ell}\in\mathbb{C}^{(2\ell+1)\times R}$ denotes the degree-$\ell$ coefficient matrix of the~$j$-th noisy observation, and write~$Y_{j,\ell}^{m,q}$ for its entry at spherical-harmonic order~$m$ and radial-basis index~$q$. For each admissible degree triple~$(\ell_1,\ell_2,\ell_3)$ satisfying the coupling condition~$|\ell_1-\ell_2|\leq \ell_3\leq \ell_1+\ell_2$, and for radial-basis indices~$(q_1,q_2,q_3)$, the empirical bispectrum entry is estimated as

\begin{equation}
	\label{eq:empirical_bispectrum_so3}
		\widehat{B}_{\ell_1,\ell_2,\ell_3}[q_1,q_2,q_3]=\frac{1}{n}\sum_{j=1}^{n} \sum_{\substack{|m_1|\leq \ell_1,\ |m_2|\leq \ell_2\\ m_1+m_2+m_3=0}} (-1)^{m_1} \Gamma_{\ell_1,\ell_2,\ell_3}^{m_1,m_2} Y_{j,\ell_1}^{m_1, q_1} Y_{j,\ell_2}^{m_2, q_2} Y_{j,\ell_3}^{m_3, q_3},
\end{equation}
where~$m_3=-m_1-m_2$, the summation is restricted to~$|m_3|\leq \ell_3$, and~$\Gamma_{\ell_1,\ell_2,\ell_3}^{m_1,m_2}$ is the corresponding coupling coefficient, given in~\cite{bendory2025orbit}.

The recovery is performed by frequency marching. The bands~$\ell=0$ and~$\ell=1$ are initialized as in Section~\ref{sec:MLE_so3}, thereby fixing the global rotational ambiguity. Suppose that all coefficient blocks of degrees smaller than~$\ell$ have already been recovered. To recover the degree-$\ell$ block, we use bispectrum equations associated with lower-degree pairs of the form~$(i,\ell-i,\ell)$, for~$i=1,\ldots,\lfloor \ell/2\rfloor$, together with the corresponding admissible permutations. In these equations, the factors at degrees~$i$ and~$\ell-i$ are already known, while the degree-$\ell$ factor is unknown. Therefore, each such bispectrum equation is linear in the coefficients~$F_{\ell,m,q}$.

Collecting these equations over the selected lower-degree pairs, Clebsch--Gordan indices, and radial triples yields a linear system for the full degree-$\ell$ coefficient block. Under the generic full-rank conditions established in~\cite{bendory2025orbit}, this system determines the degree-$\ell$ block uniquely. Repeating this step for~$\ell=2,\ldots,L$ recovers the signal up to the fixed global rotation. In our implementation, we compute only the bispectrum slices required for these marching equations, rather than materializing the full bispectrum tensor.

\subsection{Computational complexity analysis for MRA over~$\SO(3)$}
\label{sec:complexity_so3}
Recall that~$n$ denotes the number of measurements, $L$ the spherical-harmonic bandlimit, $R$ the number of  radial degrees, and $d_\ell=2\ell+1$.
For the $\SO(3)$ computations, we denote by~$N_{\Omega}^{(\mathrm{fast})}$ the number of sampled rotations used to approximate the integrals over~$\SO(3)$ in the proposed fast approximate MLE algorithm, and by~$N_{\Omega}^{(\mathrm{EM})}$ the number of rotations in the discretized rotation grid used by EM.

\subsubsection{Fast approximate MLE}
The proposed fast approximate MLE first estimates the second moments~$\widehat{M}_\ell$ in~\eqref{eq:second_moment} for all degrees~$\ell=0,\ldots,L$.    Since~$Y_{j,\ell}\in\mathbb{C}^{d_\ell\times R}$, forming all empirical second moments costs~$\mathcal{O}(nR^2\sum_{\ell=0}^{L}d_\ell)=\mathcal{O}(nR^2L^2)$.    Each~$\widehat{M}_\ell\in\mathbb{C}^{R\times R}$ is then factorized by a dense Hermitian eigendecomposition, with cost~$\mathcal{O}(LR^3)$.
The dominant cost of one likelihood-based marching sweep is approximating the integrals defining~$s_{j,\ell}[0]$ and~$s_{j,\ell}[\ell]$ in~\eqref{eq:s_so3} over a sampled rotation grid in~$\SO(3)$, and then forming the matrices~$A_\ell$.
In the implementation of Algorithm~\ref{alg:fast_mle_so3},
the exponent in the alignment kernel~$H_{j,\ell}$ in~\eqref{eq:so3_alignment_kernel} is updated incrementally on the sampled rotation grid: after degree~$\ell$ is recovered, its contribution is added to the kernel used for subsequent degrees, while unresolved higher-degree coefficient matrices remain set to zero.
Thus, each update uses the current lower-degree kernel, without recomputing the full sum over degrees from scratch.
With this implementation, the data-dependent cost of one forward marching sweep is~$\mathcal{O} \, (n(R+N_{\Omega}^{(\mathrm{fast})})L^3)$.
The factor~$L^3$ comes from summing the squared Wigner-$D$ dimensions over all degrees, since~$\sum_{\ell=0}^{L}d_\ell^2=\mathcal{O}(L^3)$.
The Procrustes updates require SVDs of matrices of size~$d_\ell\times d_\ell$, with total cost~$\sum_{\ell=0}^{L}\mathcal{O}(d_\ell^3)=\mathcal{O}(L^4)$, independent of~$n$.
Therefore, in the large-sample regime, the dominant per-sweep cost of the proposed fast approximate MLE is~$\mathcal{O}\, (n(R+N_{\Omega}^{(\mathrm{fast})})L^3)$.

\subsubsection{EM}
The E-step~\eqref{eq:e_step_so3} evaluates the likelihood weights for all~$n$ observations and all~$N_{\Omega}^{(\mathrm{EM})}$ sampled rotations. For each degree~$\ell$, applying the Wigner-$D$ matrix~$D_\ell(\omega)\in\mathbb{C}^{d_\ell\times d_\ell}$ to the coefficient matrix~$F_\ell\in\mathbb{C}^{d_\ell\times R}$ costs~$\mathcal{O}(d_\ell^2R)$. Summing over all degrees gives~$\sum_{\ell=0}^{L}d_\ell^2R=\mathcal{O}(RL^3)$. Hence, the E-step has per-iteration cost~$\mathcal{O} \, (n N_{\Omega}^{(\mathrm{EM})}RL^3)$. The M-step~\eqref{eq:m_step_so3} has the same order of dependence on~$n$, $N_{\Omega}^{(\mathrm{EM})}$, $R$, and~$L$. Therefore, for~$T_{\mathrm{EM}}$ iterations, the total complexity scales as~$\mathcal{O} \, (T_{\mathrm{EM}}n N_{\Omega}^{(\mathrm{EM})} RL^3)$. As in the~$\SO(2)$ case, the main computational drawback is the iteration count~$T_{\mathrm{EM}}$, which  grows in the low-SNR regime.

\subsubsection{Bispectrum-based invariant method}
We consider the frequency marching bispectrum implementation described in Section~\ref{sec:bispectrum_so3}. The implementation avoids materializing the full bispectrum tensor. For each target degree~$\ell$, it computes only the bispectrum slices involving one unknown degree-$\ell$ block and two already recovered lower-degree blocks. Specifically, it uses lower-degree pairs of the form~$(i,\ell-i,\ell)$, for~$i=1,\ldots,\lfloor \ell/2\rfloor$, together with the corresponding admissible permutations. The resulting bispectrum equations are linear in the coefficients of~$F_\ell$, and recovery proceeds degree by degree by solving least-squares problems. The dominant cost is the empirical construction of the required bispectrum slices. The radial triples~$(r_1,r_2,r_3)$ contribute a factor~$R^3$. For each selected degree triple, the bispectrum entry requires an inner summation over the spherical-harmonic orders~$m_1,m_2,m_3$ with~$m_1+m_2+m_3=0$. Summing this angular cost over all pairs and all target degrees gives a total factor~$\mathcal{O}(L^4)$. Thus, the dominant cost of the bispectrum baseline is~$\mathcal{O}\left(nR^3L^4\right)$. After these empirical bispectrum slices are formed, the recovery stage solves linear systems degree by degree. This stage does not scale with~$n$, with complexity depending on the number of equations retained in the least-squares systems.

\subsubsection{Comparison}
The three methods exhibit different computational trade-offs. EM directly optimizes the marginalized likelihood and is therefore potentially statistically efficient, but it is iterative: its total cost scales with the number of EM iterations~$T_{\mathrm{EM}}$, which can grow substantially in the low-SNR regime. The bispectrum method is non-iterative and avoids estimating the rotations, but in the~$\SO(3)$ setting it requires estimating and manipulating high-dimensional third-order invariant tensors. Its dominant cost is~$\mathcal{O}(nR^3L^4)$, reflecting in particular the cubic dependence on the number of radial degrees of freedom.

The proposed fast approximate MLE avoids EM iterations and replaces the global likelihood optimization by a sequence of likelihood-based marching updates.
Its dominant per-sweep cost is~$\mathcal{O} \, (n(R+N_{\Omega}^{(\mathrm{fast})})L^3)$, where~$N_{\Omega}^{(\mathrm{fast})}$ is the number of sampled rotations used to approximate the likelihood integrals over~$\SO(3)$. For functions on~$\SO(3)$ bandlimited at degree~$L$, the number of Wigner-$D$ coefficients is~$\sum_{\ell=0}^{L}(2\ell+1)^2=\mathcal{O}(L^3)$, and sampling theorems on~$\SO(3)$ use a number of samples proportional to~$L^3$~\cite{mcewen2015novel}.
Thus, a natural scaling is~$N_{\Omega}^{(\mathrm{fast})}=\Theta(L^3)$. In the proportional-shell regime~$R=\Theta(L)$, this gives a per-sweep complexity of~$\mathcal{O}(nL^6)$, compared with~$\mathcal{O}(nL^7)$ for the bispectrum baseline. The numerical experiments in the following section show that this reduced scaling leads to a favorable runtime--accuracy trade-off in the regimes considered, increasingly so as the problem dimension grows.

\section{Numerical Experiments}
\label{sec:numerical_experiments}
\begin{figure}[t]
	\centering
	\begin{subfigure}[b]{0.48\linewidth}
		\centering
		\includegraphics[width=\linewidth, keepaspectratio]{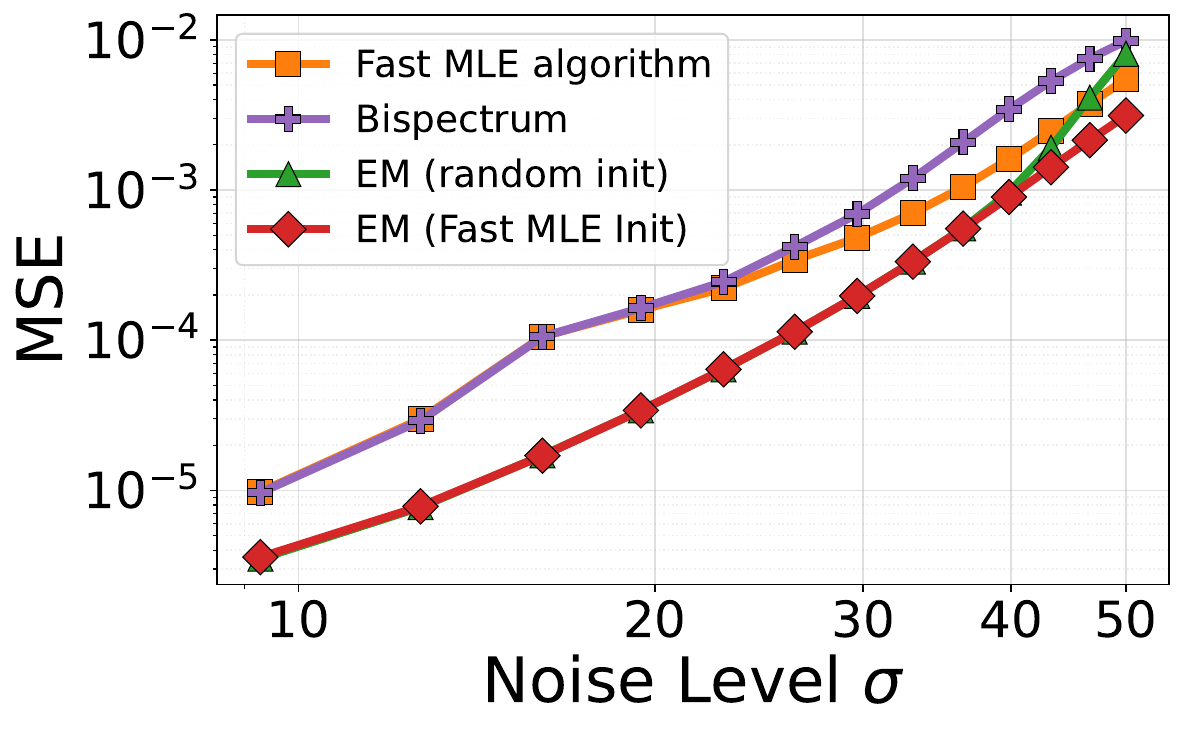}
		\caption{Error vs. $\sigma$.}
		\label{fig:noise_mse}
	\end{subfigure}
	\hfill
	\begin{subfigure}[b]{0.48\linewidth}
		\centering
		\includegraphics[width=\linewidth, keepaspectratio]{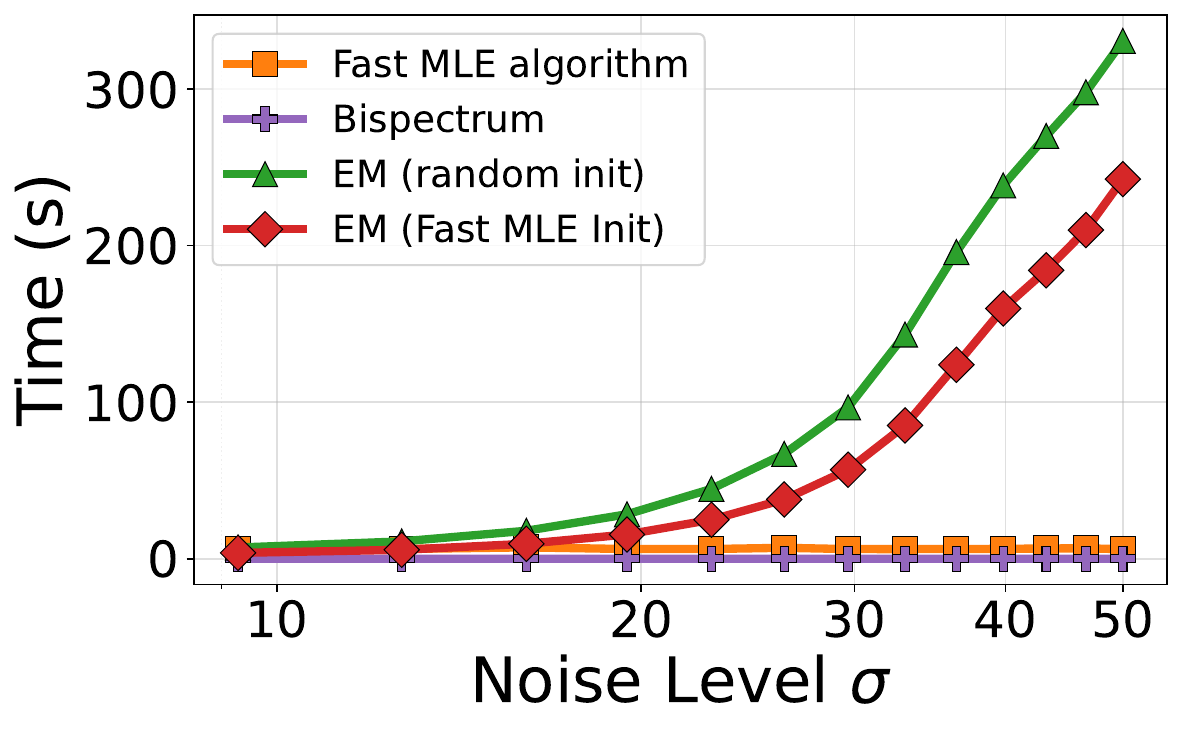}
		\caption{Runtime vs. $\sigma$.}
		\label{fig:noise_time}
	\end{subfigure}
	
	\caption{Performance comparison in the~$\SO(2)$ MRA model as a function of the noise level~$\sigma$, with~$n=100{,}000$ observations. Panel~(a) reports the reconstruction error after alignment, and panel~(b) reports the runtime. As the SNR decreases, the accuracy gap between the proposed fast approximate MLE and EM decreases. The proposed method maintains an essentially constant runtime across noise levels, while EM becomes increasingly expensive in the low-SNR regime.}
	\label{fig:so2_combined_results}
\end{figure}

\begin{figure}[t]
	\centering
		\vspace{0pt}
		\centering
		\begin{subfigure}[t]{\linewidth}
			\centering
			\includegraphics[width=0.85\linewidth, keepaspectratio]{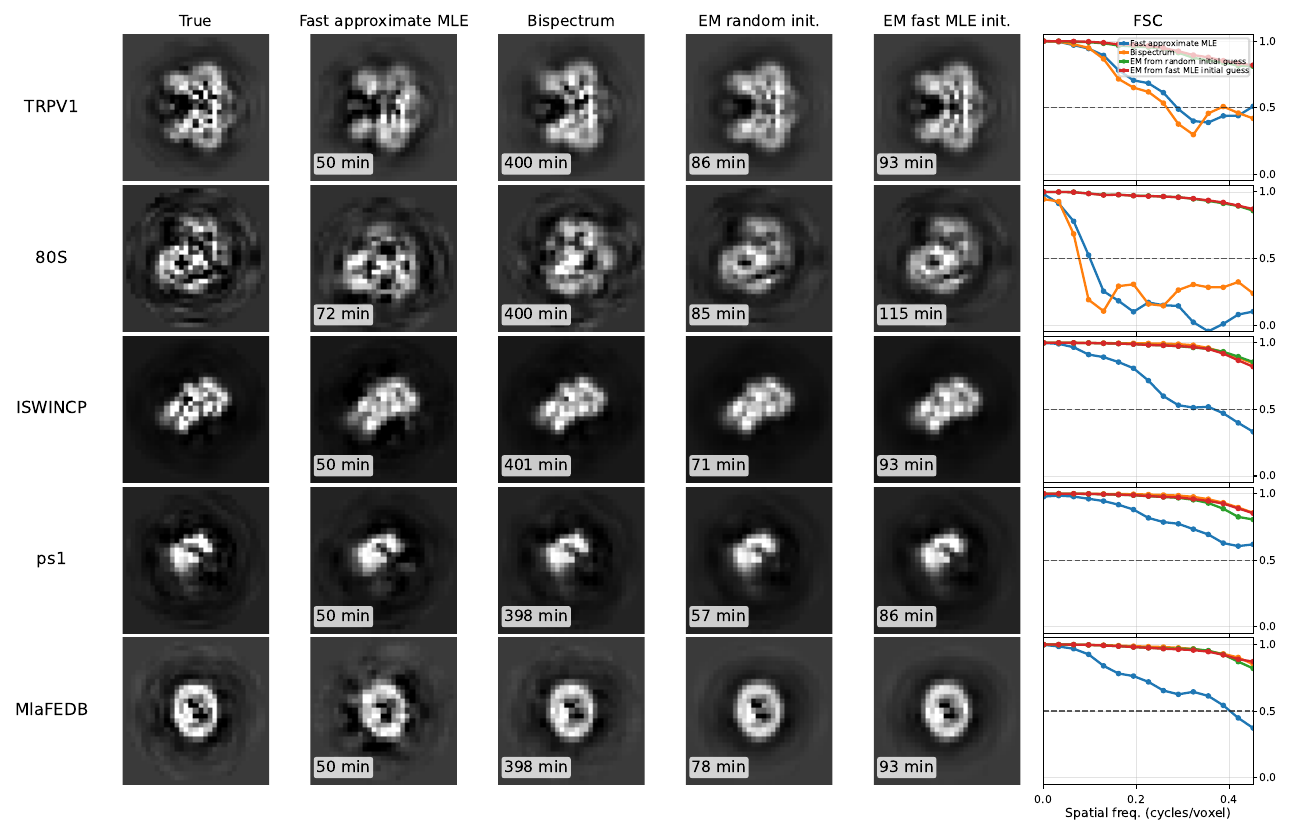}
			\caption{Five molecular volumes at~$\mathrm{SNR}=1/50$.}
			\label{fig:combined_results}
		\end{subfigure}
		\vspace{0.4em}
		\begin{subfigure}[t]{\linewidth}
			\centering
			\includegraphics[width=0.85\linewidth, keepaspectratio]{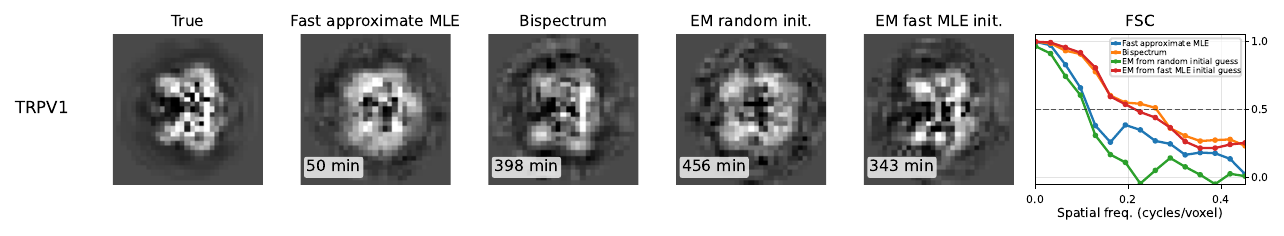}
			\caption{TRPV1 at~$\mathrm{SNR}=1/500$.}
			\label{fig:low_snr_trpv1}
		\end{subfigure}
	\hfill
	\caption{Comparison of reconstruction methods in the~$\SO(3)$ MRA model. Each row shows the central slice of the ground-truth volume and the aligned reconstructions. The reconstructed volumes are provided in the supplementary material. The number shown in each reconstruction panel is the corresponding running time in minutes; for EM initialized from fast MLE, this includes the fast-MLE initialization time. The rightmost column reports the Fourier shell correlation~(FSC) between each aligned reconstruction and the ground truth; the dashed horizontal line marks the FSC value~$0.5$. Panel~(a) compares the methods across five molecular volumes at~$\mathrm{SNR}=1/50$, while panel~(b) shows a lower-SNR TRPV1 experiment at~$\mathrm{SNR}=1/500$. }
	\label{fig:numerical_results}
\end{figure}

We evaluate the proposed fast approximate MLE algorithm in both the~$\SO(2)$ and~$\SO(3)$ MRA models. The~$\SO(2)$ experiment is designed to illustrate the low-SNR behavior predicted by the likelihood expansion: as the noise level increases, the leading terms of the marginalized likelihood become increasingly dominant, and the gap between the proposed closed-form estimator and the MLE approximation computed by EM is expected to decrease. The~$\SO(3)$ experiments constitute the main computational demonstration of the paper, showing that the same likelihood-expansion principle yields fast and useful low-frequency reconstructions of molecular volumes. In all experiments, we compare the proposed method with EM and with a bispectrum-based invariant baseline. All running times reported in this section were measured on a MacBook Pro with an Apple M3 Pro chip, comprising 11~CPU cores, and 36~GB of unified memory. For reproducibility, the implementations of the algorithms described in this paper are publicly available at~\url{https://github.com/krshay/Fast-MLE-MRA}.


We begin with the~$\SO(2)$ MRA model. 
We generated real bandlimited signals with bandlimit~$L=5$. The real and imaginary parts of the Fourier coefficients were drawn independently from a standard normal distribution, subject to the conjugate-symmetry constraint. The observations were generated according to~\eqref{eq:observations}, with rotation angles drawn uniformly from~$[0,2\pi)$ and additive white Gaussian noise with variance~$\sigma^2$. We used a discretization grid of~$N_{\theta}^{(\mathrm{EM})} = N_{\theta}^{(\mathrm{fast})} =1000$ points and a convergence tolerance of~$10^{-6}$ for EM.

Reconstruction accuracy is measured by the mean squared error of the Fourier coefficients after alignment. Since the signal is identifiable only up to a global rotation~$\alpha$, we first align the estimate to the ground truth by minimizing the~$\ell_2$ distance over a grid of shifts. The reported error is~$\mathrm{MSE} = \sum_{k=0}^{L} |\widehat{\hf}[k]e^{-ik\alpha^{\ast}}-\hf[k]|^2/\|\hf\|^2$,
where~$\widehat{\hf}$ denotes the estimated Fourier coefficients,~$\hf$ denotes the ground truth, and~$\alpha^{\ast}$ is the optimal shift.

Figures~\ref{fig:noise_mse} and~\ref{fig:noise_time} compare reconstruction error and runtime as a function of the noise level~$\sigma$, with~$n=100{,}000$ observations. The errors were averaged over~$10$ trials. As shown in Figure~\ref{fig:noise_mse}, the gap between the proposed fast approximate MLE and EM decreases as~$\sigma$ increases. This behavior is consistent with the low-SNR derivation: in the high-noise regime, the leading informative term in the marginalized likelihood dominates, and the closed-form update captures increasingly much of the information used by the full likelihood. Figure~\ref{fig:noise_time} shows that this comes with a substantial computational advantage. The runtime of the proposed method remains essentially constant as the noise level increases, whereas EM becomes slower in the low-SNR regime as analyzed in~\cite{balanov2025expectation}.


We next consider the~$\SO(3)$ MRA model. We evaluate the proposed fast approximate MLE algorithm on five molecular volumes: the TRPV1 structure~\cite{gao2016trpv1}, available from the Electron Microscopy Data Bank (EMDB) under accession code~\mbox{EMD-8117}; the 80S ribosome~\cite{wong2014cryo} (EMD-2660); ISWINCP~\cite{yan2019structures} (EMD-9718); ps1~\cite{plaschka2017structure} (EMD-3682); and MlaFEDB~\cite{coudray2020structure} (EMD-22116). For the numerical simulations, we instantiate the general radial basis in~\eqref{eq:signal_so3} with a truncated spherical-Bessel basis, so that in spherical coordinates the volume is expanded as
\begin{equation}
	X(r,\theta,\varphi) = \sum_{\ell=0}^{L}\sum_{m=-\ell}^{\ell}\sum_{q=1}^{R} F_{\ell,m,q}\,\frac{4j_\ell(\nu_{\ell q}\frac{r}{c})}{|j_{\ell+1}(\nu_{\ell q})|} \mathcal{Y}_{\ell}^ m(\theta,\varphi),
\end{equation}
where~$\mathcal{Y}_{\ell}^m$ are complex spherical harmonics,~$j_\ell$ is the spherical Bessel function of order~$\ell$,~$\nu_{\ell q}$ is the~$q$-th positive zero of~$j_\ell$, and~$c$ is the radius of the ball on which the volume is represented. In our experiments,~$c=1/2$. The angular index~$\ell$ is truncated at~$L$, and for each angular degree we retain~$R$ radial shells. The coefficients at band~$\ell$ can be collected into a matrix~$F_\ell\in\mathbb{C}^{(2\ell+1)\times R}$. 

Synthetic observations are generated according to~\eqref{eq:measurements_so3}. We define the signal-to-noise ratio as~$\mathrm{SNR} = \frac{\sum_{\ell=0}^{L}\|F_\ell\|_{\F}^{2}}{\sigma^2 \sum_{\ell=0}^{L}(2\ell+1)R}$. For each volume, we downsample the reference map to a cubic grid of size~$31^3$, normalize it, and expand it in the spherical-Bessel basis. We use angular bandlimit~$L=12$ and~$R=30$ radial shells, with~$n=10^6$ observations. 	For the likelihood-based methods, the integral over~$\SO(3)$ is approximated using~$N_{\Omega}^{(\mathrm{fast})} = N_{\Omega}^{(\mathrm{EM})} = 2500$ sampled rotations. EM is run until convergence, defined as either a maximum relative coefficient update below~$10^{-6}$ or a relative log-likelihood improvement below~$10^{-4}$.

Since the~$\SO(3)$ MRA model is identifiable only up to a global rotation, all reconstructed volumes are aligned to the reference before evaluation. The alignment is performed using the~\texttt{fitmap} command in ChimeraX~\cite{meng2023ucsf}. We compare the methods using Fourier shell correlation~(FSC), a standard resolution metric in cryo-EM. Given two volumes~$X$ and~$\widehat X$, with Fourier transforms~$\mathcal{F}X$ and~$\mathcal{F}\widehat X$, the FSC at spatial-frequency shell~$S_\rho$ is defined as
\begin{equation}
	\mathrm{FSC}(\rho)=\frac{\sum_{\xi\in S_\rho}(\mathcal{F}X)(\xi)\overline{(\mathcal{F}\widehat X)(\xi)}}{\sqrt{\sum_{\xi\in S_\rho}|(\mathcal{F}X)(\xi)|^2\sum_{\xi\in S_\rho}|(\mathcal{F}\widehat X)(\xi)|^2}}.
\end{equation}
Higher FSC values indicate better agreement with the ground-truth volume at the corresponding spatial frequency. We report the FSC curves and use the common~$0.5$ threshold as a simple summary of the spatial frequency up to which the reconstruction remains strongly correlated with the ground truth.

Figure~\ref{fig:numerical_results} illustrates the accuracy-runtime trade-offs of the four methods across moderate and very low SNR regimes. At~$\mathrm{SNR}=1/50$, the proposed fast approximate MLE provides a rapid low-frequency reconstruction across all tested molecular volumes: its running time is substantially lower than that of the bispectrum baseline, while its FSC curves remain informative at low and intermediate spatial frequencies. The bispectrum method often improves the reconstruction quality, but at a significantly higher computational cost. EM initialized from the fast MLE estimate consistently improves upon the fast MLE alone and typically attains the best FSC curves among the tested methods, while EM initialized randomly is less reliable.

The low-SNR TRPV1 experiment in Figure~\ref{fig:low_snr_trpv1} shows that the same qualitative behavior persists, and becomes more pronounced, at~$\mathrm{SNR}=1/500$. Even in this severe noise regime, the fast approximate MLE produces a meaningful low-frequency reconstruction in under one hour. Bispectrum inversion and EM initialized from the fast MLE achieve somewhat higher FSC values, but require several times more computation. In contrast, EM initialized randomly deteriorates substantially, highlighting the importance of a reliable initialization. These results support the role of the proposed estimator both as a fast standalone low-frequency reconstruction method and as an effective initialization for likelihood-based refinement in low-SNR settings.

\section{Discussion and outlook}
\label{sec:discussion}
\subsection{Likelihood-based frequency marching}
\label{sec:why_frequency_marching}
The proposed method can be viewed as a hybrid between maximum-likelihood estimation and bispectrum inversion. As in maximum-likelihood estimation, each update is derived from the marginalized log-likelihood and is therefore tied directly to the statistical model. As in bispectrum inversion, the unknown degrees of freedom are resolved by frequency marching, from low to high frequencies. The connection to the bispectrum is, however, more specific than this shared marching structure: our consistency analysis (Sections~\ref{sec:consistency_so2} and~\ref{sec:consistency_so3}) shows that, in the low-SNR limit, the leading-order term of each likelihood update is a third-order rotational invariant carrying the same cross-frequency information as the corresponding bispectrum entries. The key distinction is that the proposed method never forms or inverts these invariant tensors explicitly; the cubic interactions arise implicitly from the low-SNR expansion of the likelihood and yield closed-form band updates. This perspective also clarifies why the unresolved higher-frequency components may be set to zero during the initial forward sweep: the leading-order marching update couples each new component only to previously recovered lower ones, so the low-order likelihood information already suffices to identify the components sequentially.

More broadly, the same mechanism is likely to extend to other compact group actions whose representations decompose into ordered frequency bands, most directly~$\SO(n)$: whenever second-order invariants determine each band up to a finite-dimensional unitary ambiguity and the leading low-SNR term resolving that ambiguity is cubic, coupling each new band to previously recovered ones, one expects an analogous sequential, closed-form likelihood-based estimator. Making the precise conditions explicit is left to future work.

\subsection{A fast approximate MLE for cryo-EM}
A natural direction for future work is to extend the likelihood-expansion approach developed in this paper to the full cryo-EM model, where the observations are noisy two-dimensional tomographic projections of an unknown three-dimensional volume at unknown rotations~\cite{bendory2020single}. Although this setting includes additional complications beyond the $\SO(3)$ MRA model, most notably the projection operator, the same principle may still apply: after a suitable parametrization of the volume by unitary matrix degrees of freedom, a low-SNR expansion of the marginalized likelihood could reduce the leading-order estimation problem to a sequence of Procrustes updates. This would mirror the phase-recovery step in~\eqref{eq:solution_phase} and the matrix recovery step in~\eqref{eq:solution_Q_so3}. We leave the full derivation, together with an efficient implementation, to future work. If successful, this approach could provide a fast \emph{ab initio} estimator for cryo-EM, serving as a fast initialization for standard likelihood-based refinement pipelines~\cite{scheres2012relion,punjani2017cryosparc}.

	\section*{Acknowledgment}
	The authors thank the reviewers of an earlier version of this manuscript for their careful reading and constructive comments, which helped improve the paper.

\appendix
\section{Proof of Lemma~\ref{lem:L_so2}}
\label{app:lemma_L_so2}
\begin{proof}
	Recall the definition of $H_{j,k}(\theta)$  from~\eqref{eq:H_jk}.
	Since~$H_{j,k}(\theta) > 0$,~$s_{j,k}[0] := \int_0^{2\pi}H_{j,k}(\theta)d\theta$ is real and strictly positive.
	The $\phi$-dependent part of the $j$-th summand in~\eqref{eqn:log-likelhood-simplified} is~$\log\int_0^{2\pi} H_{j,k}(\theta) e^{\frac{r_k}{\sigma^2} \Re\left( \overline{\hy_j[k]}\phi e^{-ik\theta} \right) } d\theta$.
	Since~$\hy_j[k]=\mathcal{O}_{\mathbb{P}}(\sigma)$ in the low-SNR regime and~$r_k=\mathcal{O}(1)$, the exponent involving~$\phi$ is uniformly~$\mathcal{O}_{\mathbb{P}}(r_k/\sigma)$.
	Using~$e^x=1+x+\mathcal{O}(x^2)$ and~$\log(a+b)=\log a+b/a+\mathcal{O}(b^2/a^2)$, the expression above equals
	\begin{equation}
		\log s_{j,k}[0] + \frac{r_k}{\sigma^2} \Re\left\{ \overline{\hy_j[k]}\phi \frac{s_{j,k}[k]}{s_{j,k}[0]} \right\} + \mathcal{O}_{\mathbb{P}}\left(\frac{r_k^2}{\sigma^2}\right),
	\end{equation}
	uniformly over~$|\phi|=1$, where~$s_{j, k}[k]$ is defined in~\eqref{eq:sjk}.
	Averaging~\eqref{eqn:log-likelhood-simplified} over~$j$ and absorbing all terms independent of~$\phi$ into~$\tilde{C}_k$ gives~$\mathcal{L}_k(\phi) = \tilde{C}_k + \frac{r_k}{\sigma^2} \Re\left\{ \overline{Z_k}\phi \right\} +  R_k(\phi)$,	where~$Z_k$ is defined in~\eqref{eq:Zk_so2} and~$R_k(\phi)=\mathcal{O}_{\mathbb{P}}(r_k^2/\sigma^2)$ uniformly over~$|\phi|=1$.
\end{proof}

\section{Proof of Theorem~\ref{thm:consistency_so2}}    \label{app:proof_consistency_so2}
\begin{proof}
	We first record the expansion underlying one frequency-marching step. Fix $k\in \{2,\ldots,L\}$, and suppose that the coefficients at frequencies $0,\ldots,k-1$ are equal to their population values, while the unresolved higher frequencies are set to zero.
	Define
	\begin{equation}
		\widetilde{H}_{j,k}(\theta) \triangleq e^{ \frac{1}{\sigma^2} \sum_{m=0}^{k-1} \Re\left\{\overline{\hy_j[m]}\hf[m]e^{-{i}m\theta} \right\} },
		\label{eqn:tilde_H_def}
	\end{equation}
	and
	\begin{equation}
		\widetilde{\mu}_{j,k} \triangleq \frac{\int_0^{2\pi} e^{i k \theta} \widetilde{H}_{j,k}(\theta) d\theta} {\int_0^{2\pi} \widetilde{H}_{j,k}(\theta) d\theta}, \quad \widetilde{Z}_k \triangleq \frac{1}{n}\sum_{j=1}^n \hy_j[k] \widetilde{\mu}_{j,k}.
	\end{equation}
	Let $\mathcal{P}_k=\{(a,b)\in \{1,\ldots,k-1\}^2:a+b=k\}$.
	We record the following lemma.
	
	\begin{lemma}    \label{lem:oracle_update}
		For each fixed $k\in\{2,\ldots,L\}$, 
		\begin{equation}
			\widetilde{Z}_k = \frac{\alpha_k}{\sigma^4}\hf[k] + o_{\mathbb{P}}(\sigma^{-4}) + \mathcal{O}_{\mathbb{P}}\left(\frac{1}{\sigma\sqrt{n}}\right), \label{eq:oracle_Zk}
		\end{equation}
		where $\alpha_k = \frac{1}{8} \sum_{(a,b)\in\mathcal{P}_k} r_a^2 r_b^2 >0$.\footnote{For a stochastic sequence $X_n$ and a deterministic positive sequence~$a_n$, the notation~$X_n=o_{\mathbb{P}}(a_n)$ means that~$X_n/a_n$ converges to zero in probability as~$n\to\infty$.}
	\end{lemma}
	
	\begin{proof}[Proof of Lemma~\ref{lem:oracle_update}]
		We write~$\widetilde{H}_{j,k} (\theta) = e^{\widetilde{\eta}_{j,k}(\theta)}$, such that
		\begin{equation}
			\widetilde{\eta}_{j,k}(\theta) = \frac{1}{2\sigma^2} \sum_{m=0}^{k-1} \left( \overline{\hy_j[m]}\hf[m]e^{-{i}m\theta} + \hy_j[m]\overline{\hf[m]}e^{{i}m\theta} \right). \label{eq:oracle_score_expanded_short}
		\end{equation}
		Since $L$ and $\hf$ are fixed, each coefficient~$\hy_j[m]$ is~$\mathcal{O}_{\mathbb{P}}(\sigma)$ in the low-SNR regime, and therefore~$\sup_{\theta\in[0,2\pi)}|\widetilde{\eta}_{j,k}(\theta)|=\mathcal{O}_{\mathbb{P}}(\sigma^{-1})$. The numerator of $\widetilde{\mu}_{j,k}$ is the~$(-k)$-th Fourier coefficient of~$\widetilde{H}_{j,k}$. Looking at the Taylor expansion of~$e^{\widetilde{\eta}_{j,k}}$, we identify that the constant term in the expansion has only frequency zero, and the linear term~$\widetilde{\eta}_{j,k}$ contains only the frequencies~$0,\pm1,\ldots,\pm(k-1)$, and therefore they do not interact with~$e^{ik\theta}$. The first possible contribution is therefore the quadratic term~$\widetilde{\eta}_{j,k}^2/2$. Indeed, multiplying two frequency terms gives~$e^{i a\theta}e^{i b\theta}=e^{{i}(a+b)\theta}$, so the mode~$e^{-i k\theta}$ appears precisely from pairs~$(a,b) \in \mathcal{P}_k$. Thus, the remaining terms in the expansion are given by~$\frac{1}{8\sigma^4} \sum_{(a,b)\in\mathcal{P}_k} \overline{\hy_j[a]} \,\overline{\hy_j[b]}\,\hf[a]\hf[b]$. The denominator of~$\widetilde{\mu}_{j,k}$ satisfies~$\int_0^{2\pi}e^{\widetilde{\eta}_{j,k}(\theta)}d\theta = 2\pi\left(1+\mathcal{O}_{\mathbb{P}}(\sigma^{-2})\right)$ because the integral of the linear term vanishes. Thus,
		\begin{equation}
			\widetilde{\mu}_{j,k} = \frac{1}{8\sigma^4} \sum_{(a,b)\in\mathcal{P}_k} \overline{\hy_j[a]} \, \overline{\hy_j[b]} \,\hf[a]\hf[b] + R_{j,k}, \label{eq:mu_expansion_short}
		\end{equation}
		where $R_{j,k}$ collects the higher-order terms in the expansion.
		Substituting~\eqref{eq:mu_expansion_short} into the definition of $\widetilde{Z}_k$ yields
		\begin{equation}
			\widetilde{Z}_k = \frac{1}{8n\sigma^4} \sum_{j=1}^n \sum_{(a,b)\in\mathcal{P}_k} \hy_j[k] \, \overline{\hy_j[a]}\, \overline{\hy_j[b]}\hf[a]\hf[b] + \frac{1}{n}\sum_{j=1}^n \hy_j[k]R_{j,k}. \label{eq:Zk_cubic_short}
		\end{equation}
		For each $(a,b)\in\mathcal{P}_k$, the deterministic rotation phases in the leading cubic statistic cancel because $a+b=k$. Therefore, we have that~$\mathbb{E}\left[ \hy_j[k] \, \overline{\hy_j[a]} \, \overline{\hy_j[b]} \right] = \hf[k]\, \overline{\hf[a]}\, \overline{\hf[b]}$, and the expectation of the leading term in~\eqref{eq:Zk_cubic_short} is
		\begin{equation}
			\frac{1}{8\sigma^4} \sum_{(a,b)\in\mathcal{P}_k} r_a^2 r_b^2\hf[k] = \frac{\alpha_k}{\sigma^4}\hf[k].
		\end{equation}
		The higher-order Taylor terms, together with the correction from the denominator, contribute $o_{\mathbb{P}}(\sigma^{-4})$ after multiplication by $\hy_j[k]$ and averaging over the observations. Moreover, the leading cubic summand in~\eqref{eq:Zk_cubic_short} has standard deviation of order $\sigma^{-1}$ after the prefactor $\sigma^{-4}$ is applied, and therefore its empirical fluctuation is $\mathcal{O}_{\mathbb{P}}(1/(\sigma\sqrt{n}))$.
	\end{proof}
	
	We now prove the theorem.
	The preliminary estimators satisfy $\widehat{\hf}[0]\xrightarrow{p}\hf[0]$, and $\widehat{r}_k\xrightarrow{p}|\hf[k]|$, for $k=1,\ldots,L$, because $n=\omega(\sigma^6)$ implies the weaker requirements $n=\omega(\sigma^2)$ for the mean and $n=\omega(\sigma^4)$ for the second moments.
	The phase of the first coefficient is fixed by the gauge, so~$\widehat{\hf}[1]=\widehat{r}_1\xrightarrow{p}\hf[1]$.
	
	Assume inductively that~$\widehat{\hf}[m]\xrightarrow{p}\hf[m]$ for all~$m<k$. The expansion in Lemma~\ref{lem:oracle_update} is stable under consistent perturbations of the lower-frequency coefficients, since the Taylor expansion is uniform in a fixed neighborhood of the population coefficients and~$L$ is fixed. Therefore, the actual update statistic~$\widehat{Z}_k$ computed by Algorithm~\ref{alg:fast_mle} satisfies the same expansion as~\eqref{eq:oracle_Zk}, with~$\widetilde{Z}_k$ replaced by $\widehat{Z}_k$. Since~$n=\omega(\sigma^6)$, multiplication by $\sigma^4$ gives $\sigma^4\widehat{Z}_k\xrightarrow{p}\alpha_k\hf[k]$. As $\alpha_k>0$ and $\hf[k]\neq0$, this implies~$\widehat{Z}_k/|\widehat{Z}_k|\xrightarrow{p}\hf[k]/|\hf[k]|$. Combining this phase consistency with~$\widehat{r}_k\xrightarrow{p}|\hf[k]|$ gives~$\widehat{\hf}[k]\xrightarrow{p}\hf[k]$.
\end{proof}
\begin{remark}[Sample complexity of the single-pass variant]
	\label{rem:single_pass_rate}
	For the single-pass kernel~$H_j^{(1)}$ of Remark~\ref{rem:single_pass_variant}, frequency~$k$ first appears at the~$k$-th order Taylor term of~$e^{\eta_j^{(1)}}$, rather than at the second order as in Lemma~\ref{lem:oracle_update}. The same argument then gives $\widetilde Z_k^{(1)}=\beta_k\sigma^{-2k}\hf[k] + o_{\mathbb{P}}(\sigma^{-2k}) + \mathcal{O}_{\mathbb{P}}(\sigma^{1-k}/\sqrt n)$ for some~$\beta_k>0$, so consistency at frequency~$k$ requires~$n=\omega(\sigma^{2k+2})$. Taking the maximum over~$k=2,\ldots,L$, the single-pass estimator is consistent only under~$n=\omega(\sigma^{2L+2})$, compared with the marching algorithm's uniform~$n=\omega(\sigma^6)$.
\end{remark}

\section{Proof of Lemma~\ref{lem:L_so3}}
\label{app:lemma_L_so3}
\begin{proof}
	For fixed~$j$ and~$\ell$, the $Q$-dependent part of the $j$-th summand in~\eqref{eq:L_so3_restricted} is~$\log \int_{\SO(3)} H_{j,\ell}(\omega) e^{x_{j,\ell}(\omega;Q) } d\omega$, where
	\begin{equation}
		x_{j,\ell}(\omega;Q) := \frac{1}{\sigma^2} \Re\tr\left( \widetilde{F}_\ell Y_{j,\ell}^{*}D_\ell(\omega) Q \right).
	\end{equation}    Since~$\|Y_{j,\ell}\|_{\F}=\mathcal{O}_{\mathbb{P}}(\sigma)$ as $\sigma \to \infty$  and~$\|\widetilde{F}_\ell\|_{\F}=\mathcal{O}(1)$, uniformly over~$Q\in\U(2\ell+1)$ and~$\omega\in\SO(3)$ we have $x_{j,\ell}(\omega;Q) = \mathcal{O}_{\mathbb{P}} ( \|\widetilde{F}_\ell\|_{\F}/\sigma)$.
	Using~$e^x=1+x+\mathcal{O}(x^2)$ and~$\log(a+b)=\log a+b/a+\mathcal{O}(b^2/a^2)$, the expression above equals
	\begin{equation}
		\log s_{j,\ell}[0] + \frac{1}{\sigma^2} \frac{ \Re\tr\left( \widetilde{F}_\ell Y_{j,\ell}^{*}s_{j,\ell}[\ell]Q \right) }{ s_{j,\ell}[0]} + \mathcal{O}_{\mathbb{P}} \left(\frac{\|\widetilde{F}_\ell\|_{\F}^{2}}{\sigma^2} \right),
	\end{equation}
	uniformly over~$Q\in\U(2\ell+1)$, where~$s_{j, \ell}[\ell]$ is defined in~\eqref{eq:s_so3}.
	Here~$s_{j,\ell}[0]=\int_{\SO(3)}H_{j,\ell}(\omega)d\omega$ is real and strictly positive because~$H_{j,\ell}(\omega)>0$. Averaging~\eqref{eq:L_so3_restricted} over~$j=1,\ldots,n$ and absorbing all terms independent of~$Q$ into~$\tilde{C}_\ell$ gives~$\mathcal{L}_\ell(Q) = \tilde{C}_\ell + \frac{1}{\sigma^2} \Re\tr\left( A_\ell Q \right) + R_\ell(Q)$, where~$A_\ell$ is defined in~\eqref{eq:Aell_so3} and~$R_\ell(Q)=\mathcal{O}_{\mathbb{P}}(\|\widetilde{F}_\ell\|_{\F}^{2}/\sigma^2)$ uniformly over~$Q\in\U(2\ell+1)$.
\end{proof}

\section{Proof of Theorem~\ref{thm:consistency_so3}}
\label{app:proof_consistency_so3}
\begin{proof}
	Recall that the second-moment factorization determines each coefficient
	matrix~$F_\ell$ up to a left-acting unitary matrix~$Q_\ell\in\U(2\ell+1)$.
	The global~$\SO(3)$ ambiguity is removed by the prescribed choice of~$Q_1$,
	and the $O(3)$ component ambiguity is handled as described in
	Remark~\ref{rem:o3_component_ambiguity}. Thus, it is enough to prove that
	the marching procedure consistently recovers the matrices~$Q_\ell$ in this
	fixed gauge. The argument follows the same structure as the~$\SO(2)$ proof
	(Appendix~\ref{app:proof_consistency_so2}), with the scalar phases replaced
	by the unitary matrices~$Q_\ell$.
	
	The preliminary estimators satisfy~$\widehat{F}_0\xrightarrow{p}F_0$, and
	$\widetilde{F}_\ell\xrightarrow{p}\Fpop_\ell$ for~$\ell=1,\ldots,L$,
	because $n=\omega(\sigma^6)$ implies the weaker
	requirements~$n=\omega(\sigma^2)$ for the mean and~$n=\omega(\sigma^4)$
	for the second moments, and~$F_\ell$ is full row rank. Consequently the
	empirical row metric $\Sigma_\ell:=\widetilde F_\ell\widetilde F_\ell^{*}$
	from~\eqref{eq:tilde_F} satisfies
	$\Sigma_\ell\xrightarrow{p}\breve\Sigma_\ell:=\Fpop_\ell\Fpop_\ell^{*}$.
	The degree~$\ell=1$ is fixed by the global gauge, so the induction starts
	from~$\ell=1$.
	
	We first record the expansion underlying one degree-marching step, at the
	oracle level where the lower bands are fixed at their population values.
	Fix $\ell\in\{2,\ldots,L\}$, and suppose that the coefficient matrices at
	degrees $0,\ldots,\ell-1$ equal~$F_0,\ldots,F_{\ell-1}$, while all
	unresolved higher-degree matrices are set to zero. Define the alignment
	weight
	\begin{equation}
		\label{eq:H_jl}
		\widetilde{H}_{j,\ell}(\omega) \triangleq
		e^{ \frac{1}{\sigma^2}\sum_{q=0}^{\ell-1}
			\Re\tr\left(F_q Y_{j,q}^{\ast}D_q(\omega)\right) }.
	\end{equation}
	Let
	\begin{equation}
		\begin{split}
			&\widetilde{\mu}_{j,\ell} \triangleq \frac{\int_{\SO(3)}\widetilde{H}_{j,\ell}(\omega)D_\ell(\omega)\,d\omega} {\int_{\SO(3)}\widetilde{H}_{j,\ell}(\omega)\,d\omega}, \\ &\widetilde{A}_\ell \triangleq \frac{1}{n}\,\Fpop_\ell\sum_{j=1}^n Y_{j,\ell}^{\ast} \widetilde{\mu}_{j,\ell}, \qquad \bar A_\ell := \Sigma_\ell^{\dagger}\widetilde A_\ell,
		\end{split}
	\end{equation}
	where~$\bar A_\ell$ is the oracle counterpart of the normalized matrix~\eqref{eq:Aell_whitened_so3}. Let $\mathcal{P}_\ell=\{(q_1,q_2)\in\{0,\ldots,\ell-1\}^2:
	|q_1-q_2|\leq\ell\leq q_1+q_2\}$.
	We record the following lemma.
	
	\begin{lemma}[Degree-update expansion over~$\SO(3)$]
		\label{lem:oracle_update_so3}
		For each fixed $\ell\in\{2,\ldots,L\}$,
		\begin{equation}
			\bar{A}_\ell
			= \frac{1}{\sigma^4}\,\check\Gamma_\ell\,Q_\ell^{*}
			+ o_{\mathbb{P}}(\sigma^{-4})
			+ \mathcal{O}_{\mathbb{P}}\!\left(\frac{1}{\sigma\sqrt{n}}\right),
			\label{eq:oracle_Aell_expansion}
		\end{equation}
		where $\check\Gamma_\ell\in\mathbb{C}^{(2\ell+1)\times(2\ell+1)}$ is the
		deterministic degree-$\ell$ coupling defined in~\eqref{eq:checkGamma_def},
		Hermitian positive definite under the genericity and full-row-rank
		assumptions on $\{F_q\}_{q\le\ell}$.
	\end{lemma}
	
	\begin{proof}[Proof of Lemma~\ref{lem:oracle_update_so3}]
		We write~$\widetilde{H}_{j,\ell}(\omega)=e^{\widetilde{\eta}_{j,\ell}(\omega)}$,
		such that
		\begin{equation}
			\widetilde{\eta}_{j,\ell}(\omega)
			= \frac{1}{\sigma^2}\sum_{q=0}^{\ell-1}
			\Re\tr\left(F_q Y_{j,q}^{\ast}D_q(\omega)\right).
			\label{eq:eta_so3_oracle}
		\end{equation}
		Since~$L$,~$R$, and the signal are fixed, the entries of~$Y_{j,q}$ are
		$\mathcal{O}_{\mathbb{P}}(\sigma)$ as~$\sigma\to\infty$, and thus
		$\sup_{\omega\in\SO(3)}|\widetilde{\eta}_{j,\ell}(\omega)|
		=\mathcal{O}_{\mathbb{P}}(\sigma^{-1})$. The numerator of
		$\widetilde{\mu}_{j,\ell}$ is the~$\ell$-th Fourier coefficient
		over~$\SO(3)$ of $\widetilde{H}_{j,\ell}$. In the Taylor expansion
		of~$e^{\widetilde{\eta}_{j,\ell}}$, the constant term contains only the
		trivial representation, and the linear term~$\widetilde{\eta}_{j,\ell}$
		contains only the degrees~$0,\ldots,\ell-1$; neither interacts
		with~$D_\ell(\omega)$.
		
		The role of the identity~$a+b=k$ in the~$\SO(2)$ proof is now played by
		the Clebsch--Gordan product rule. Products of two Wigner-$D$ matrices of
		degrees $q_1$ and~$q_2$ decompose into irreducible
		degrees~$r=|q_1-q_2|,\ldots,q_1+q_2$, and thus a product of two
		lower-degree terms can contain degree~$\ell$ precisely when
		$|q_1-q_2|\leq\ell\leq q_1+q_2$~\cite{bendory2025orbit}. Therefore the
		degree-$\ell$ component of the quadratic
		term~$\widetilde{\eta}_{j,\ell}^2/2$ is obtained by coupling the
		admissible lower-degree pairs $(q_1,q_2)\in\mathcal{P}_\ell$ through the
		Clebsch--Gordan coefficients. The denominator satisfies
		$\int_{\SO(3)}e^{\widetilde{\eta}_{j,\ell}(\omega)}\,d\omega
		=1+\mathcal{O}_{\mathbb{P}}(\sigma^{-2})$ because the Haar integral of
		the nontrivial linear components vanishes. Thus,
		\begin{equation}
			\widetilde{\mu}_{j,\ell}
			= \sigma^{-4}\sum_{(q_1,q_2)\in\mathcal{P}_\ell}
			\Gamma_{j,\ell}^{(q_1,q_2)} + R_{j,\ell},
			\label{eq:Sjell_expansion_so3}
		\end{equation}
		where~$\Gamma_{j,\ell}^{(q_1,q_2)}$ denotes the degree-$\ell$
		Clebsch--Gordan coupling of the degree-$q_1$ and degree-$q_2$ factors,
		and~$R_{j,\ell}$ collects the higher-order Taylor terms and the
		denominator correction. Substituting into~$\widetilde{A}_\ell$ shows
		that the leading update is cubic in the observations:
		\begin{equation}
			\widetilde{A}_\ell
			= \frac{1}{n\sigma^4}\,\Fpop_\ell\sum_{j=1}^n
			\sum_{(q_1,q_2)\in\mathcal{P}_\ell}
			Y_{j,\ell}^{\ast}\Gamma_{j,\ell}^{(q_1,q_2)}
			+ \frac{1}{n}\,\Fpop_\ell\sum_{j=1}^n Y_{j,\ell}^{\ast}R_{j,\ell}.
			\label{eq:Aell_cubic_so3}
		\end{equation}
		
		For each $(q_1,q_2)\in\mathcal{P}_\ell$, the latent rotation in the
		leading cubic statistic cancels: by equivariance of the
		Clebsch--Gordan coupling, the signal part
		of~$\Gamma_{j,\ell}^{(q_1,q_2)}$ carries the
		factor~$D_\ell(\omega_j)$, while~$Y_{j,\ell}^{\ast}$
		contributes~$F_\ell^{\ast}D_\ell(\omega_j)^{\ast}$; since~$D_\ell(\omega_j)$
		is unitary, $D_\ell(\omega_j)^{\ast}D_\ell(\omega_j)=I$ and the rotation
		drops out. Therefore
		$\mathbb{E}\left[Y_{j,\ell}^{\ast}\,\Gamma_{j,\ell}^{(q_1,q_2)}\right]
		= F_\ell^{\ast}\,\Gamma_\ell^{(q_1,q_2)}$, where~$\Gamma_\ell^{(q_1,q_2)}$
		is the corresponding deterministic coupling of the true lower-degree
		coefficients. Using $\Fpop_\ell=Q_\ell^{*}F_\ell$ and
		$\Fpop_\ell F_\ell^{*}=\Fpop_\ell\Fpop_\ell^{*}Q_\ell^{*}
		=\breve\Sigma_\ell Q_\ell^{*}$, the expectation of the leading term
		in~\eqref{eq:Aell_cubic_so3} is
		\begin{equation}
			\frac{1}{\sigma^4} \Fpop_\ell F_\ell^{*} \sum_{(q_1,q_2)\in\mathcal P_\ell} \Gamma_\ell^{(q_1,q_2)} = \frac{1}{\sigma^4}\,\breve\Sigma_\ell\,\check\Gamma_\ell\,Q_\ell^{*},
			\label{eq:Aell_pop_unwhitened}
		\end{equation}
		where
		\begin{equation}
			\check\Gamma_\ell
			:= \sum_{(q_1,q_2)\in\mathcal P_\ell}
			Q_\ell^{*}\,\Gamma_\ell^{(q_1,q_2)}\,Q_\ell
			\label{eq:checkGamma_def}
		\end{equation}
		is the true coupling conjugated into the frame of~$\Fpop_\ell$
		by the unitary~$Q_\ell$. Multiplying~\eqref{eq:Aell_pop_unwhitened}
		on the left by~$\breve\Sigma_\ell^{\dagger}$ removes the row metric and
		gives the population limit $\sigma^{-4}\check\Gamma_\ell Q_\ell^{*}$
		of~$\bar{A}_\ell$. The higher-order Taylor terms, together with the
		correction from the denominator, contribute $o_{\mathbb{P}}(\sigma^{-4})$
		after multiplication by~$Y_{j,\ell}^{\ast}$ and averaging over the
		observations. Moreover, the leading cubic summand
		in~\eqref{eq:Aell_cubic_so3} has standard deviation of
		order~$\sigma^{-1}$ after the prefactor~$\sigma^{-4}$ is applied, and,
		since whitening by the deterministic $\breve\Sigma_\ell^{\dagger}$ does
		not affect the order, its empirical fluctuation
		is~$\mathcal{O}_{\mathbb{P}}(1/(\sigma\sqrt{n}))$.
		
		Each term in~\eqref{eq:checkGamma_def} is a degree-$\ell$ projection of
		a positive-semidefinite matrix, so~$\check\Gamma_\ell\succeq0$; it is
		positive definite since the degree-$\ell$ couplings span the full
		degree-$\ell$ space for generic full-row-rank lower
		bands~\cite{bendory2025orbit}.
	\end{proof}
	
	The lemma is stated for the oracle statistic~$\bar A_\ell$, in which the lower bands enter~$\widetilde H_{j,\ell}$ at their true population values~$F_0,\ldots,F_{\ell-1}$. The algorithm instead forms the alignment kernel from the recovered bands~$\widehat F_0,\ldots,\widehat F_{\ell-1}$, giving the algorithmic statistic~$\widehat{\bar A}_\ell := \Sigma_\ell^{\dagger}\widehat A_\ell$, where~$\widehat A_\ell$ is~$\widetilde A_\ell$ with~$\widehat H_{j,\ell}$ in place of~$\widetilde H_{j,\ell}$. The Taylor expansion of~$e^{\widetilde\eta_{j,\ell}}$ and the Clebsch--Gordan projections used in the proof above are uniform on a fixed neighborhood of the population lower-degree coefficients, so the map from the lower bands to~$\bar A_\ell$ is continuous there. By the inductive hypothesis~$\widehat F_q\xrightarrow{p}F_q$ for~$q<\ell$, the recovered bands lie in such a neighborhood with high probability; combined with $\Sigma_\ell\xrightarrow{p}\breve\Sigma_\ell$, this gives~$\widehat{\bar A}_\ell-\bar A_\ell=o_{\mathbb{P}}(\sigma^{-4})$, so $\widehat{\bar A}_\ell$ satisfies the same expansion~\eqref{eq:oracle_Aell_expansion} as~$\bar A_\ell$.
	
	We now close the induction. Assume~$\widehat{F}_q\xrightarrow{p}F_q$ for all~$q<\ell$. By the preceding paragraph, $\sigma^4\widehat{\bar A}_\ell\xrightarrow{p}\check\Gamma_\ell Q_\ell^{*}$ since $n=\omega(\sigma^6)$. The update for~$Q_\ell$ is the unitary Procrustes problem~$\widehat{Q}_\ell\in\argmax_{Q\in\U(2\ell+1)} \Re\tr(\widehat{\bar A}_\ell Q)$, with population objective~$Q\mapsto\Re\tr(\check\Gamma_\ell Q_\ell^{*}Q)$. Since $\check\Gamma_\ell\succ0$ is Hermitian positive definite, the unique maximizer over~$\U(2\ell+1)$ is~$Q_\ell$, and by continuity of the Procrustes map $\widehat{Q}_\ell\xrightarrow{p}Q_\ell$. Combining this with~$\widetilde{F}_\ell\xrightarrow{p}\Fpop_\ell$ gives~$\widehat{F}_\ell=\widehat{Q}_\ell\widetilde{F}_\ell\xrightarrow{p} Q_\ell\Fpop_\ell=F_\ell$. The induction closes for all~$\ell=1,\ldots,L$, and since~$L$ and~$R$ are fixed, coefficientwise convergence implies~$\widehat{F}_\ell\xrightarrow{p}F_\ell$ for all~$\ell=0,\ldots,L$ in the chosen global gauge.
\end{proof}
	
	\bibliographystyle{plain}
	\bibliography{refs}
	
\end{document}